\documentclass[journal ]{new-aiaa}
\usepackage[utf8]{inputenc}
\usepackage{textcomp}

\usepackage{color}
\usepackage{graphicx}
\usepackage{multirow}
\usepackage{amsmath}
\usepackage[version=4]{mhchem}
\usepackage{siunitx}
\usepackage{longtable,tabularx}
\usepackage{lineno}

\title{Effects of wave parameters on load reduction performance for amphibious aircrafts with V-hydrofoil}

\author{Yujin Lu\footnote{PhD candidate, College of Aerospace Engineering, 29 Yudao Street.}} \author{Shuanghou Deng\footnote{Associate Professor, College of Aerospace Engineering, 29 Yudao Street.}}
\author{Yuanhang Chen\footnote{Master candidate, College of Aerospace Engineering, 29 Yudao Street.}}
\author{Tianhang Xiao\footnote{Associate Professor, College of Aerospace Engineering, 29 Yudao Street.}}
\author{Jichang Chen\footnote{Postdoctor, College of Aerospace Engineering, 29 Yudao Street.}}
\affil{Nanjing University of Aeronautics and Astronautics, Nanjing, 210016, People’s Republic of China}

\author{Fan Liu\footnote{Researcher}}
\affil{China Aircraft Strength Research Institute, Xi'an, 710065, People’s Republic of China}

\author{Sichen Song\footnote{Master candidate, College of Aerospace Engineering, 29 Yudao Street.}}
\affil{Nanjing University of Aeronautics and Astronautics, Nanjing, 210016, People’s Republic of China}

\author{Bin Wu\footnote{Researcher, Key Aviation Scientific and Technological Laboratory of High-Speed Hydrodynamic.}}
\affil{China Special Vehicle Research Institute, Jingmen, 448035, People’s Republic of China}

\begin{document}

\maketitle

\begin{abstract}
An investigation of the influence of the hydrofoil on load reduction performance during an amphibious aircraft landing on still and wavy water is conducted by solving the Unsteady Reynolds-Averaged Navier-Stokes equations coupled with the standard $k-\omega$ turbulence model in this paper. During the simulations, the numerical wave tank is realized by using the velocity-inlet boundary wave maker coupled with damping wave elimination technique on the outlet, while the volume of fluid model is employed to track the water-air interface. Subsequently, the effects of geometric parameters of hydrofoil have been first discussed on still water, which indicates the primary factor influencing the load reduction is the static load coefficient of hydrofoil. Furthermore, the effects of descent velocity, wave length and wave height on load reduction are comprehensively investigated. The results show that the vertical load reduces more than 55$\%$ at the early stage of landing on the still water through assembling the hydrofoil for different descent velocity cases. Meanwhile, for the amphibious aircraft with high forward velocity, the bottom of the fuselage will come into close contact with the first wave when landing on crest position, and then the forebody will impact the next wave surface with extreme force. In this circumstance, the load reduction rate decreases to around 30$\%$, which will entail a further decline with the increase of wave length or wave height.

\end{abstract}

\section*{Nomenclature}


{\renewcommand\arraystretch{1.0}
\noindent\begin{longtable*}{@{}l @{\quad=\quad} l@{}}

$C_{\Delta0}$ & static load coefficient\\
$\chi_0, \psi$ & sweep and dihedral angle, $^\circ$\\
$b_0, b_1$ & root and tip chord length, m\\
$\mu$ & load reduction rate\\
$\alpha_\mathrm{w}$ & volume fraction of water\\
$L$ & length of the fuselage, m\\
$\eta, H$ & wave elevation and height, m\\
$u, w$ & horizontal and vertical velocity of wave particle, m/s\\
$T$ & wave period, s\\
$d$ & water depth, m\\
$a_z, a_x$ & vertical and horizontal acceleration, $\mathrm{m/s^2}$\\
$\upsilon_{x0}, \upsilon_{z0}$ & initial horizontal and descent velocity, m/s\\
$\varepsilon$ & ratio of wave length to fuselage length\\
$\theta$ & pitching angle, $^\circ$\\
$\sigma$ & wave steepness\\
\end{longtable*}}

\section{Introduction}
\lettrine{A}{mphibious} aircraft is a unique type of aircraft that is able to take off and land both on water and conventional airports \citep{yang2015survey,qiu2013efficient}. Due to its operational environment, amphibious aircraft may experience significant impact forces and pressure distribution during water landings, which can damage components of aircraft and endanger personnel safety. This is especially true when landing on wavy water where the shape and energy of the waves can lead to further damage on the structural integrity. Thus, practical hydrodynamic load reduction techniques for amphibious aircrafts are considered necessary and deserve to be developed.

Hydro-skis and hydrofoils are the most effective approaches to provide acceptable load alleviation and can be regarded as a means of absorbing energy by generating additional hydrodynamic force that reduces the water impact loading on the body. Extensive seaplane hydro-skis investigations, such as configuration, hydrodynamic, performance, stability and control characteristics, have been carried out by NACA \citep{pepper1966survey,pepper1968survey}. Taking the Grumman JRF-5 amphibian as an example, models with tandem hydro-skis \citep{wadlin1950tank}, single hydro-ski \citep{ramsen1951tank}, and twin hydro-skis \citep{ramsen1952tank} were discussed in detail. Adequate longitudinal stability, satisfactory landing behavior, good load reduction performance on wave were observed for the most test matrix. The load reducing performance of hydro-ski for amphibious aircraft was further studied numerically and experimentally \cite{gao2016influencing,gao2017study}, including hydro-ski impacting attitude angle, horizontal and vertical velocity of aircraft, length-width ratio, static load factor and deadrise angle of hydro-ski. The results showed that the rate of load reduction increases with the length-width ratio of the hydro-ski and that the rate could exceed 30$\%$ by reducing the static load factor with the deadrise angle varying from 15$^\circ$ to 20$^\circ$. As for hydrofoils, Benson et al. \cite{benson1943preliminary} carried out systematic investigations on the stability characteristics of several arrangements of hydrofoils proposed for use on seaplanes and high-speed surface boats, such as flat, V-shaped and curved hydrofoil tandem system using constant velocity or acceleration. Some of the important types of instability that have to be emphasized in applications of hydrofoils to seaplanes, such as monoplane system and flat hydrofoil. \textcolor{black}{A systematical computational design framework was developed by Seth et al. \cite{seth2021amphibious} to assess the performance and effectiveness of hydrofoils for amphibious aircraft water-takeoff, including configuration selections and sizing methods for hydrofoils. The position, span, and incidence angle of the hydrofoil were
optimized for minimum water-takeoff distance with consideration for the longitudinal stability of the aircraft.}

Furthermore, a large number of studies on seaplanes throughout the entire operation process, including take-off, landing, skiing, and other scenarios, have been conducted in recent years. Qiu et al. \cite{qiu2013efficient} proposed a decoupled algorithm to investigate the kinematic characteristics for taking off, whereby the aerodynamic forces of the full configuration and the hydrodynamic forces of the hull body were computed separately. The whole process was divided into a number of small time-step, and the forces were calculated at each time step. \textcolor{black}{The wave resistance of amphibious aircraft planing in waves was investigated by Zhou et al. \cite{zhou2023research}, concerning the effect of different wave elements and planing speeds on the motion response, and the conditions for the occurrence of jump motion and stability of the aircraft. A certain regularity was observed on the heave and pitch motion responses of amphibious aircraft with varying wave conditions. Particularly, when the wavelength reaches 1.38 to 2.76 times the length of the fuselage, the aircraft is more likely to turn into jumping motions with high speeds and large wave steepness.} An unstable oscillation phenomenon that may threaten the longitudinal stability of amphibious aircrafts, so-called porpoising motion, was evaluated by Duan et al. \cite{duan2019numerical}. Both slipstream caused by the propeller and external forces, viz. thrust and elevator forces, were also taken into consideration. Results highlighted the important role played by the hydrodynamic force in the heaving and pitching oscillations, while the aerodynamic forces have only a marginal effect. Like water landings of amphibious aircraft, ditching events of other aircraft types also involve similar fluid dynamics phenomena and have been widely studied numerically. The effects of initial pitching angle \citep{xiao2017development,guo2013effect,qu2016numerical,zheng2021numerical}, initial velocities \citep{lu2022on,ning2023numerical}, wave conditions \citep{woodgate2019simulation,xiao2021hydrodynamic,zhao2020numerical,chen2022numerical} on the kinematic characteristics and fluid dynamics phenomena have attracted most of the attention. From the perspective of engineering applications, the discussions about the application of hydrofoil for amphibious aircraft landing are still not documented sufficiently.

The present study aims to numerically simulate amphibious aircraft landings on still and wavy water, with and without a single hydrofoil, and to investigate the load reduction performance of the hydrofoil under different conditions. Three design parameters of the hydrofoil are first evaluated to obtain a hydrodynamically acceptable configuration, including static load coefficient $C_{\Delta0}$, sweep angle $\chi_0$ and dihedral angle $\psi$. Particular attention is paid at the effects of descent velocity, wave length and wave steepness. The present work is organized as follows. Section \ref{sec:method} presents the methodology for the numerical approaches, and describes the models and the computational setup; the main results are reported and discussed in Sec.~\ref{sec:results}; final conclusions are drawn in Sec.~\ref{sec:conclusion}.

\section{Methodology and Computational Setup}
\label{sec:method}

\subsection{Numerical Method}
The amphibious aircraft landing event is herein numerically investigated by using the commercial package Star CCM+ as the two-phase flow solver. In the present study the unsteady incompressible Reynolds-averaged Navier-Stokes equations with a standard $k-\omega$ two-equation turbulence model are solved by a finite volume method. The governing equations for the continuity condition and the momentum conservation condition can be written as:

\begin{equation}
\label{eq:continuity}
\dfrac{\partial u_i}{\partial x_i} = 0
\end{equation}

\begin{equation}
\label{eq:momentumConser}
\rho \dfrac{\partial u_i}{\partial t} + \rho \dfrac{\partial}{\partial x_j} (u_i u_j) = 
-\dfrac{\partial p}{\partial x_i} + \dfrac{\partial}{\partial x_j} (\nu \dfrac{\partial u_i}{\partial x_j} - \rho \overline{u'_i u'_j}) + \rho g_i
\end{equation}
where $u_i$ and $u_j$ ($i,j$ = 1,2,3) are the time averaged value of velocity, $x_i$ and $x_j$ ($i,j$ = 1,2,3) are the spatial coordinate components, $\rho$ the fluid density, $p$ the fluid pressure, $\nu$ the fluid kinematic viscosity, $- \rho \overline{u'_i u'_j}$ the Reynolds stress, and $g_i$ is the gravitational acceleration in the $i$-th direction.

The Semi-Implicit Pressure Linked Equations (SIMPLE) algorithm is employed to achieve an implicit coupling between pressure and velocity, and the gradient is reconstructed with the Green-Gauss Node Based method. The modified High Resolution Interface Capturing (HRIC) scheme is adopted for volume fraction transport. The convection terms and the diffusion terms are discretized by using second-order upwind and second-order central methods, respectively. The unsteady terms are discretized in the time domain by applying a second-order implicit scheme.

In order to capture the water-air interface, volume of fluid (VOF) method \citep{hirt1981volume} is adopted by introducing a phase volume fraction function $\alpha_q$ to represent the ratio of the volume of $q^{th}$ phase to the volume of the corresponding cell. In this paper, only air and water are considered. $\alpha_\mathrm{w}$, called the volume fraction of the water in the computational cell, which varies between 0 (air) and 1 (water) and is defined as:
\begin{equation}
\label{eq:alphaw}
\alpha_\mathrm{w}=V_\mathrm{w}/V,
\end{equation}
where $V_\mathrm{w}$ is the volume of water in the cell and $V$ is the volume of the cell. The volume fraction of the air in a cell can be computed as:
\begin{equation}
\label{eq:alphawa}
\alpha_\mathrm{a}=1-\alpha_\mathrm{w}.
\end{equation}
The volume of fraction is governed by the following equation:
\begin{equation}
\label{eq:alphawaTransp}
\dfrac{\partial \alpha}{\partial t} + u_i \dfrac{\partial \alpha}{\partial x_i} = 0.
\end{equation}
The effective value $\varphi_\mathrm{m}$ of any physical properties, such as density, viscosity, etc., of the mixture of water and air in the transport equations is determined by:
\begin{equation}
\label{eq:varphi}
\varphi_\mathrm{m}=\varphi_\mathrm{w}\alpha_\mathrm{w}+\varphi_\mathrm{a}(1-\alpha_\mathrm{w}).
\end{equation}

To accurately capture the dynamic behavior and the load generated by the water-entry process, the motion of the body caused by the fluid forces and moments on the surface is determined via a six degree-of-freedom (6DOF) model. The 6DOF model solves the equations for the rotation and translation of the center of mass of the object. The equation for the translation is formulated in the global inertial coordinate system:
\begin{equation}
\label{eq:Ftranslation}
M \cdot \dfrac{\mathrm{d} \boldsymbol\upsilon}{\mathrm{d}t}=\boldsymbol{F},
\end{equation}
and the rotation of the object is solved in the body local coordinate system by:
\begin{equation}
\label{eq:Mrotation}
\boldsymbol{L} \dfrac{\mathrm{d} \boldsymbol\omega}{\mathrm{d}t}+ \boldsymbol\omega \times \boldsymbol{L} \boldsymbol\omega=\boldsymbol{M}.
\end{equation}

Subsequently, a dynamic mesh strategy \citep{xiao2021hydrodynamic}, which moves the entire mesh rigidly along with the object at each time step according to the solution of the 6DOF model, is employed to deal with the relative motion between the fluid and the rigid body with single grid domain. The details about the dynamic mesh method can be found in our previous study \citep{lu2022on}.

\subsection{Velocity-inlet Wave Maker}
\begin{figure}[hbt!]
\centering
\includegraphics[width=0.5 \textwidth]{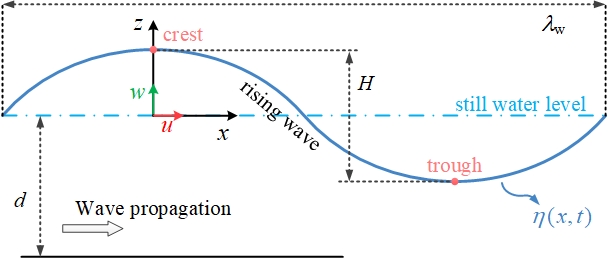}
\caption{The illustration of wave parameters and positions.}
\label{fig:wavepara}
\end{figure}

The velocity-inlet wave maker \citep{elhanafi2017effect,chen2022numerical} is used in the present simulation to generate waves within the computational domain by applying the wave elevation equation and wave velocity on the velocity-inlet boundary. In order to reduce the wave reflection at the outlet boundary, the damping wave method is employed within a specific region from the outlet boundary \citep{lin1999internal}. The wavy water is modeled by a linear wave (see Fig.~\ref{fig:wavepara}) and the elevation of the wave described herein is computed according to the linear wavy theory \citep{craik2004the} as follows:
\begin{equation}
\eta = {\dfrac{H}{2}} \cos(kx- \omega t)
\end{equation}

\noindent where $\eta$ is the wave elevation, $H$ is the wave height, $k$ is wave number, $\omega$ is the angular wave frequency. The horizontal velocity $u$ and vertical velocity $w$ of the wave can be further obtained as:
\begin{equation}
\left\{
        \begin{array}{lr}
        u={\dfrac{\pi H}{T}} \cdot {\dfrac{\cosh k(z+d)}{\sinh kd}} \cdot \cos(kx- \omega t) &  \vspace{1ex} \\
        w={\dfrac{\pi H}{T}} \cdot {\dfrac{\sinh k(z+d)}{\sinh kd}} \cdot \sin(kx- \omega t) &  
        \end{array}
\right.
\label{eq:WaveVelocity}
\end{equation}

\noindent where $T$ is the wave period, $d$ is the water depth. According to the previous work \citep{xiao2021hydrodynamic}, the grid distribution along the $x$-direction and $z$-direction are 80 grids per wave length and 20 grids per wave height which was verified to be fine enough to calculate the wave accurately..

\subsection{Models and Computational Setup}
\begin{figure}[hbt!]
\centering
\includegraphics[width=0.6 \textwidth]{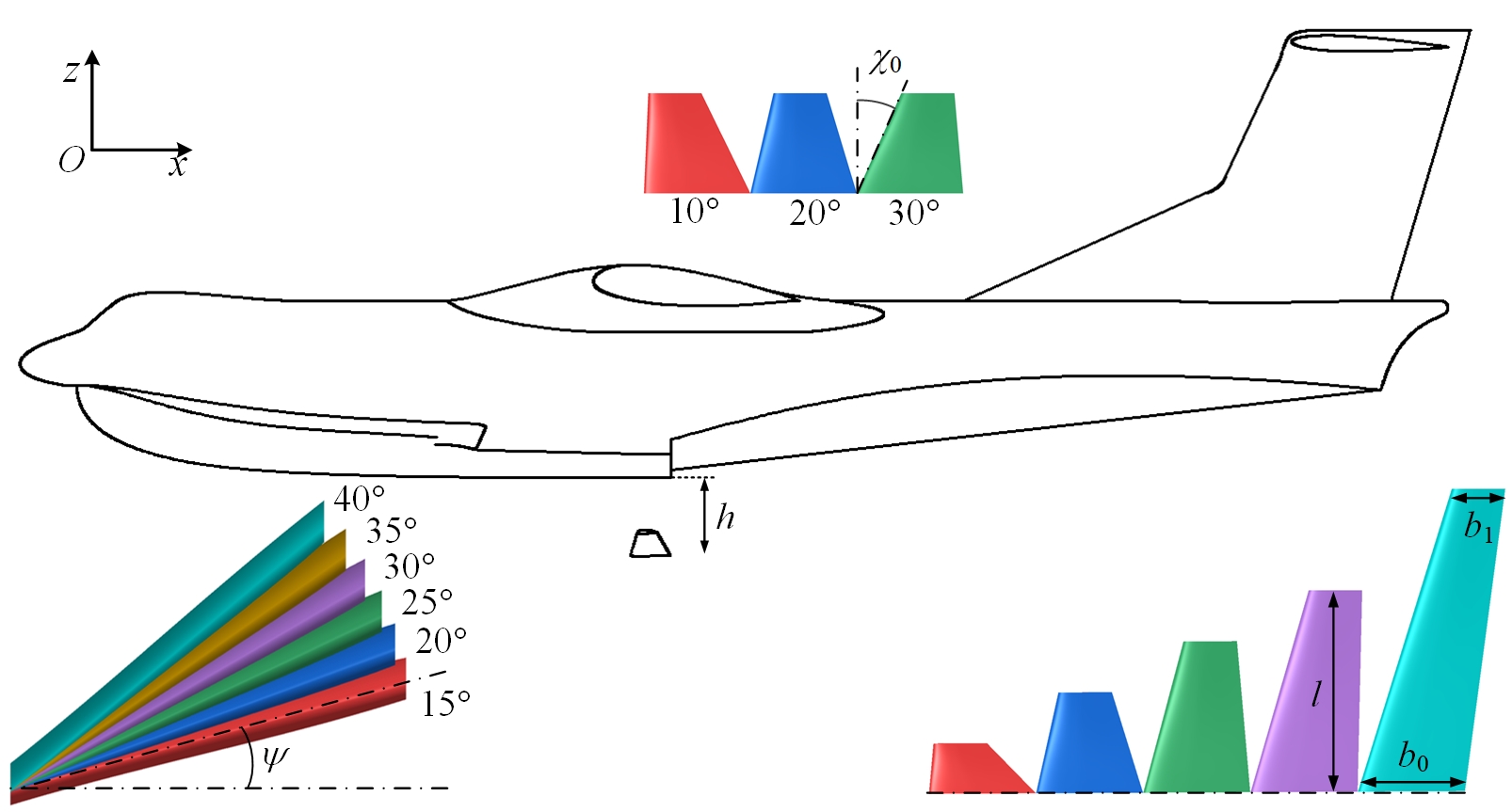}
\caption{Models and parameters of hydrofoil and aircraft}
\label{fig:model}
\end{figure}

Fig.~\ref{fig:model} shows the side view of an amphibious aircraft and the relative position of the hydrofoil with three key design parameters, such as length $l$, sweep angle $\chi_0$ (leading edge) and dihedral angle $\psi$. Note that in the present study root chord $b_0$ and tip chord $b_1$ are two constant values, 1m and 0.5m respectively, and the height of struts $h$ is 2m. Moreover, a non-dimensional variable static load coefficient $C_{\Delta0}$ \citep{wadlin1944preliminary} is introduced here to examine the effect of hydrofoil length in the following discussion and it can be expressed as:

\begin{equation}
C_{\Delta0} = \dfrac{M}{\rho_\mathrm{w} \cdot l^3}
\end{equation}

\begin{figure}[hbt!]
\centering
\includegraphics[width=.49\textwidth]{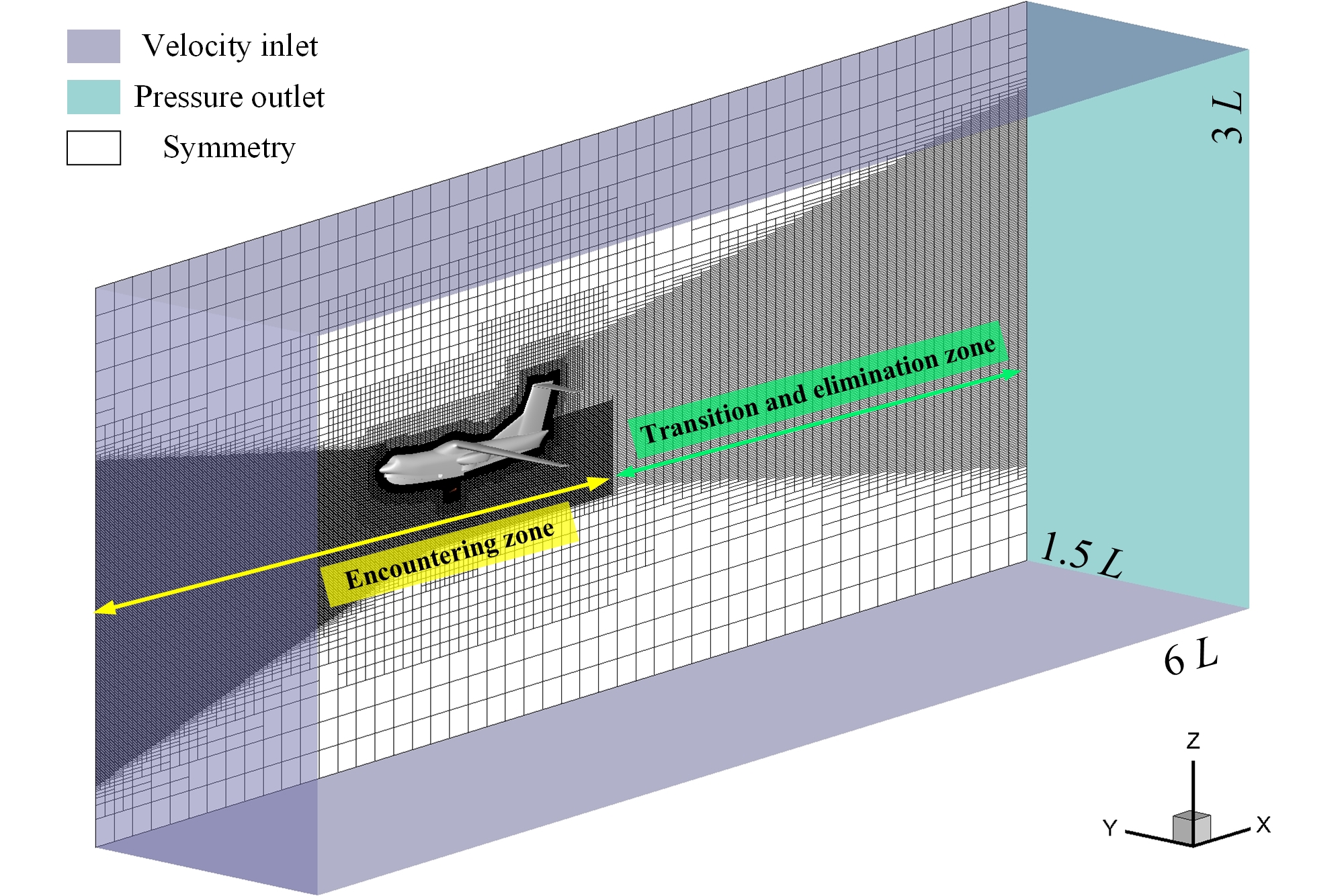}
\includegraphics[width=.49\textwidth]{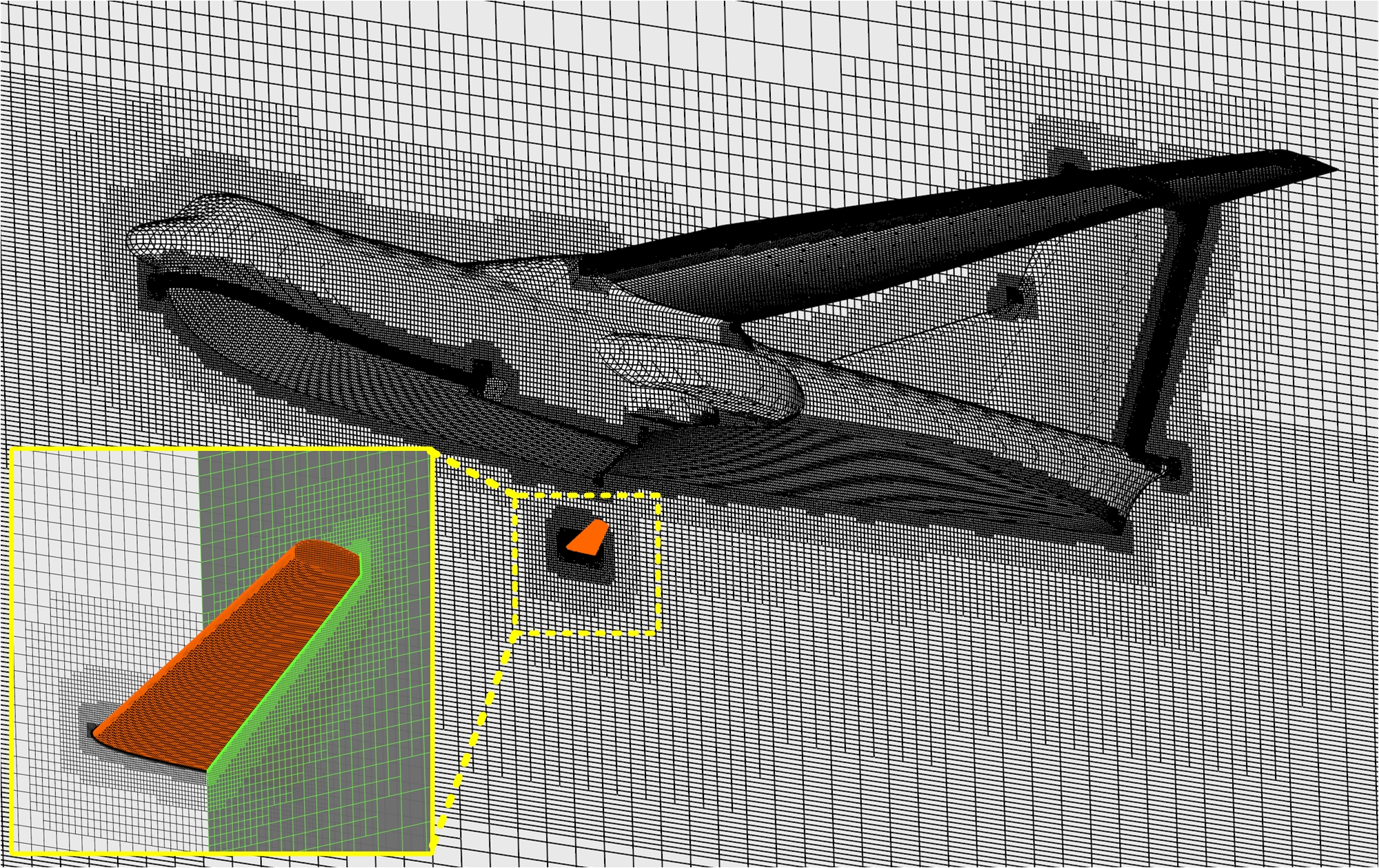}\\
\textbf{a)}\qquad\qquad\qquad\qquad\qquad\qquad\qquad\qquad\qquad\qquad\qquad\textbf{b)}
\caption{Computational domain and grid distribution: a) the size of domain; b) zoom-in mesh of aircraft and hydrofoil.}
\label{fig:domain}
\end{figure}

The semi-model and meshes used in the present study are shown in Fig.~\ref{fig:domain}. Note that only three degrees of freedom, i.e., pitch, horizontal and vertical movement, are accounted for in the present simulation. The computational domain was created by a cuboid with a size of $6L \times 1.5L \times 3L$ ($L$ is the length of the fuselage), and is regarded as large enough for the present study, as shown in Fig.~\ref{fig:domain}a. The whole domain was discretized with Cartesian cells and prismatic boundary layer grids surrounding the model and moving rigidly without deforming. Moreover, two levels for refining meshes were assigned to the entire domain as follows: encountering zone for the accurate description of the hydrodynamic features around the body and wave; transition and wave elimination zone to enable the large range of pitch motion and less wave reflection. The cell height is governed by the corresponding wave profile and the total number of grid cells in the whole domain is nearly 10 million.

\subsection{Validation}

\begin{figure}[hbt!]
\centering
\includegraphics[width=.4\textwidth]{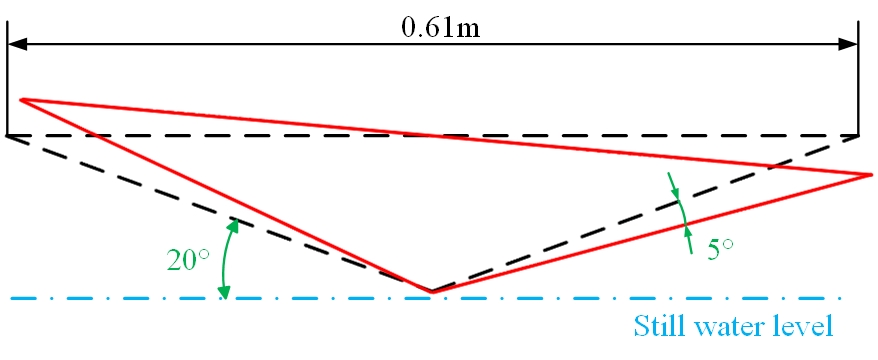}
\includegraphics[width=.4\textwidth]{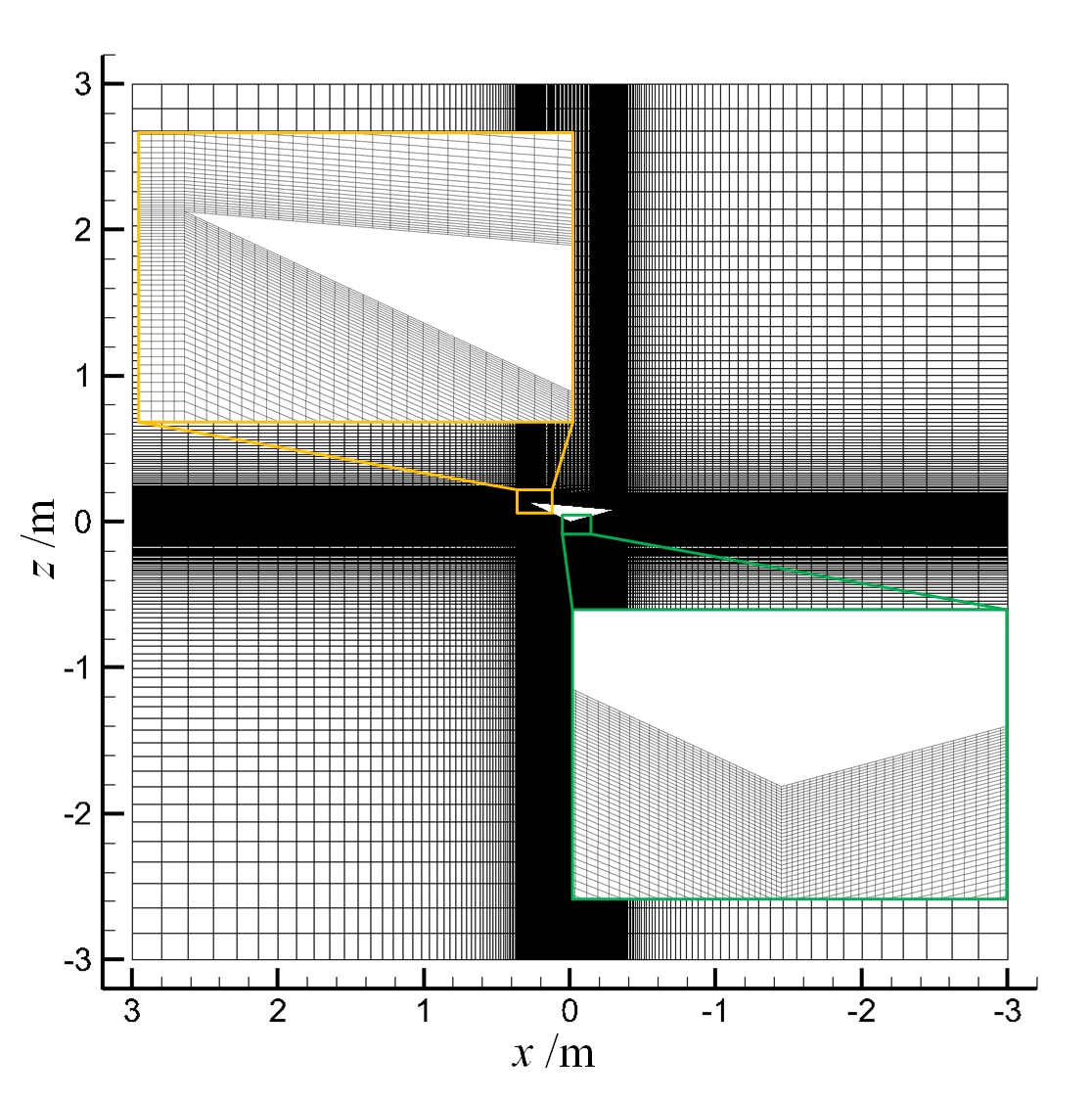}\\
\textbf{a)}\qquad\qquad\qquad\qquad\qquad\qquad\qquad\qquad\qquad\textbf{b)}
\caption{Geometry and grid topology of wedge.}
\label{fig:Wedge}
\end{figure}

Since both of the hydrofoil and fuselage of the amphibious aircraft have a V-shaped cross section, the accuracy and efficiency of the numerical method have been first validated for the free-falling water entry of a V-shaped section wedge with a 5$^\circ$ heel angle  (see Fig.~\ref{fig:Wedge}a). The simulated wedge has a width $W$ =0.61 m and a deadrise angle $\beta$ =20$^{\circ}$ referring to the work of Xu et al. \cite{xu1999asymmetric}. Fig.~\ref{fig:Wedge}b shows the mesh topology and grid density with two zoom-in views. The length of the square boundary is ten times the width of the wedge and the computational domain is discretized with structured quadrilateral grids. Inlet velocity conditions are enforced at the right, left and bottom boundaries, whereas pressure boundary condition is specified at the top boundary.

\begin{figure}[hbt!]
\centering
\includegraphics[width= 0.65\textwidth]{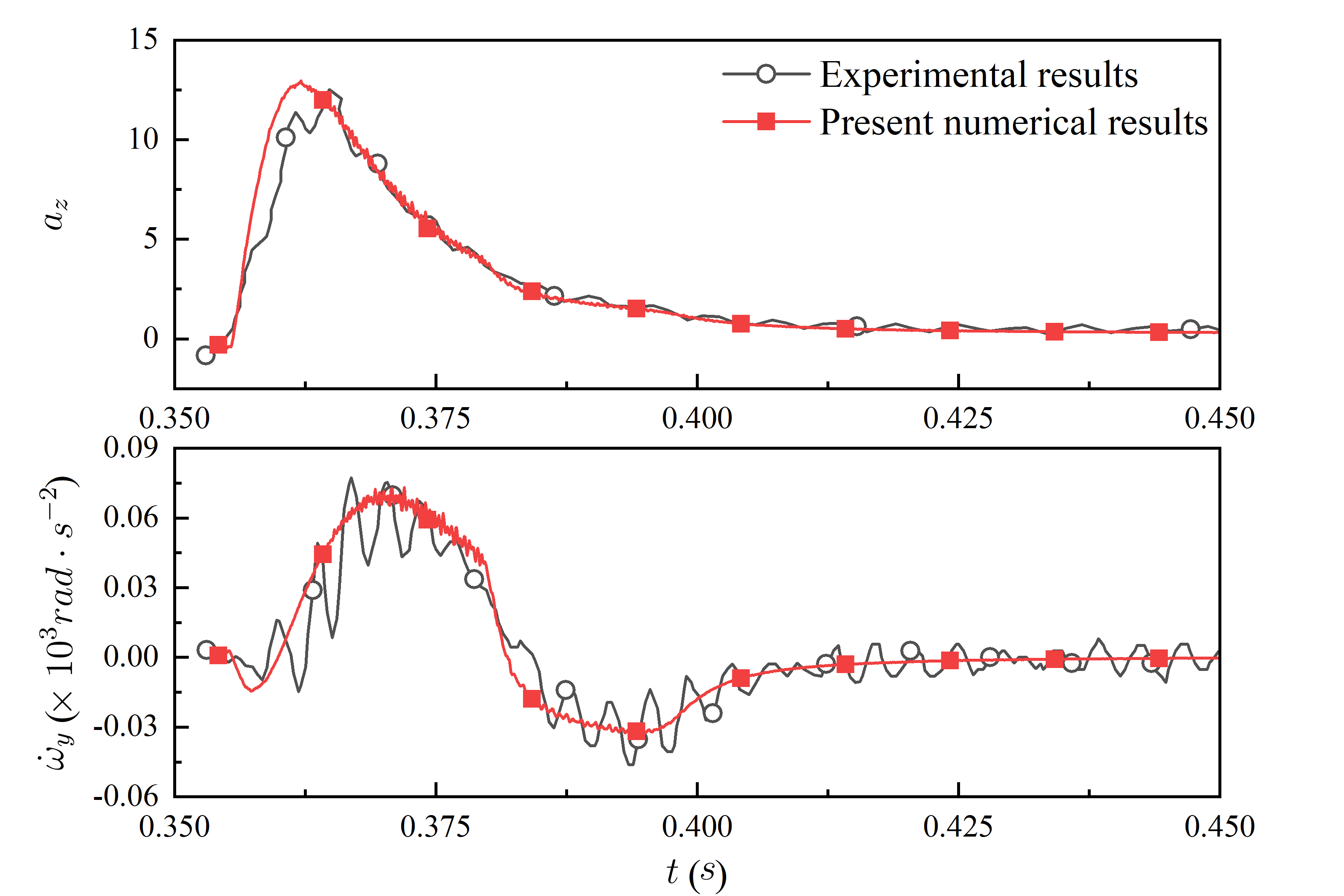}
\caption{Comparison among the present study and experimental data on the vertical water-entry of a wedge: a) dimentionless vertical acceleration $a_z$; b) angular acceleration $\Dot{\omega}_y$.}
\label{fig:WedgeExpNum}
\end{figure}

Fig.~\ref{fig:WedgeExpNum} shows the comparison between the numerical results of the present study and experimental data \citep{xu1999asymmetric} in terms of non-dimensional acceleration $a_z$ and angular acceleration $\Dot{\omega}_y$. It can be seen that the numerical results are in good agreement with experimental data, although only three degrees of freedom are considered in the present study compared to six degrees of freedom in the experiments. Overall, numerical results exhibit an overall satisfactory agreement with the experimental data.

\begin{figure}[hbt!]
\centering
\includegraphics[width=.49\textwidth]{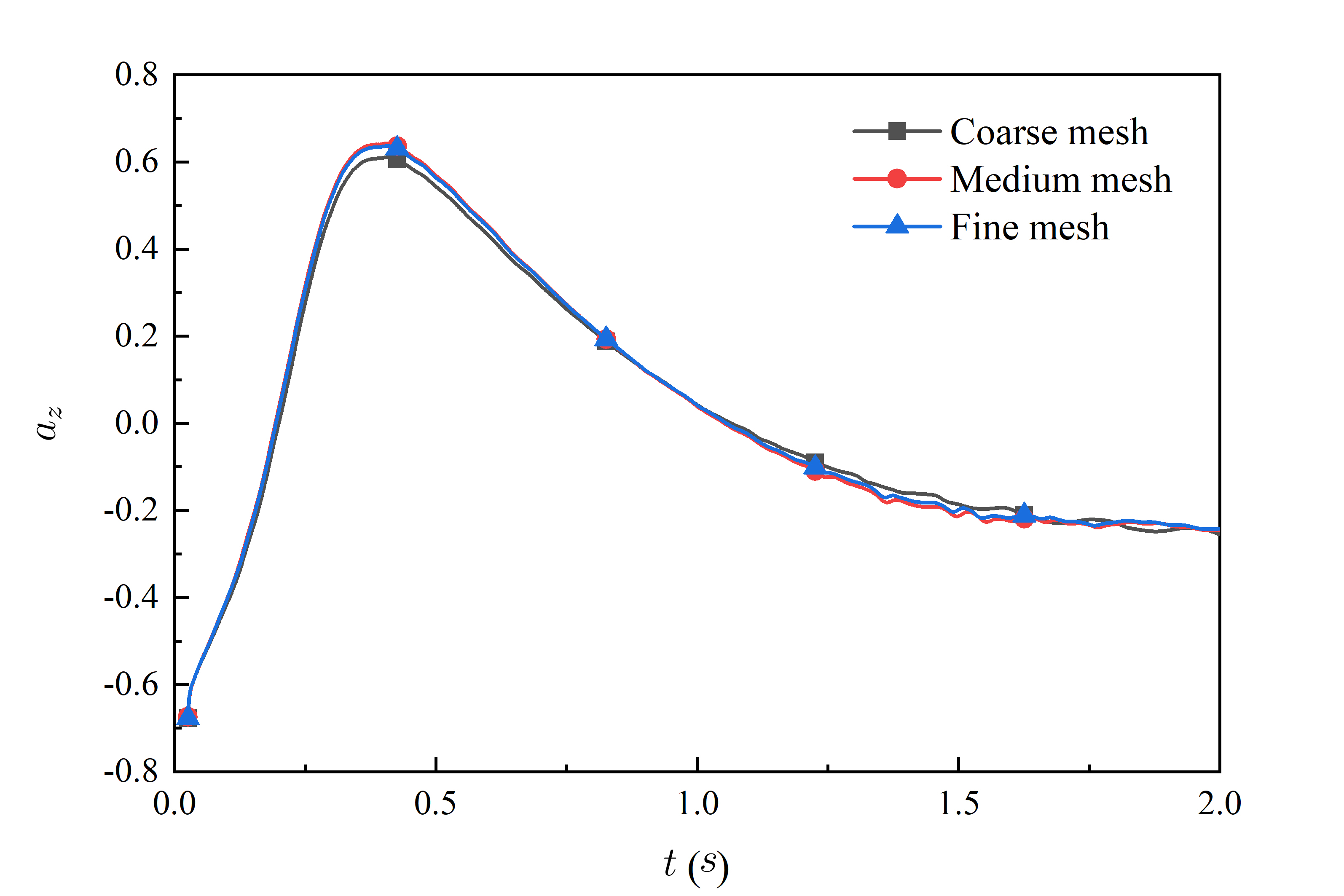}
\includegraphics[width=.49\textwidth]{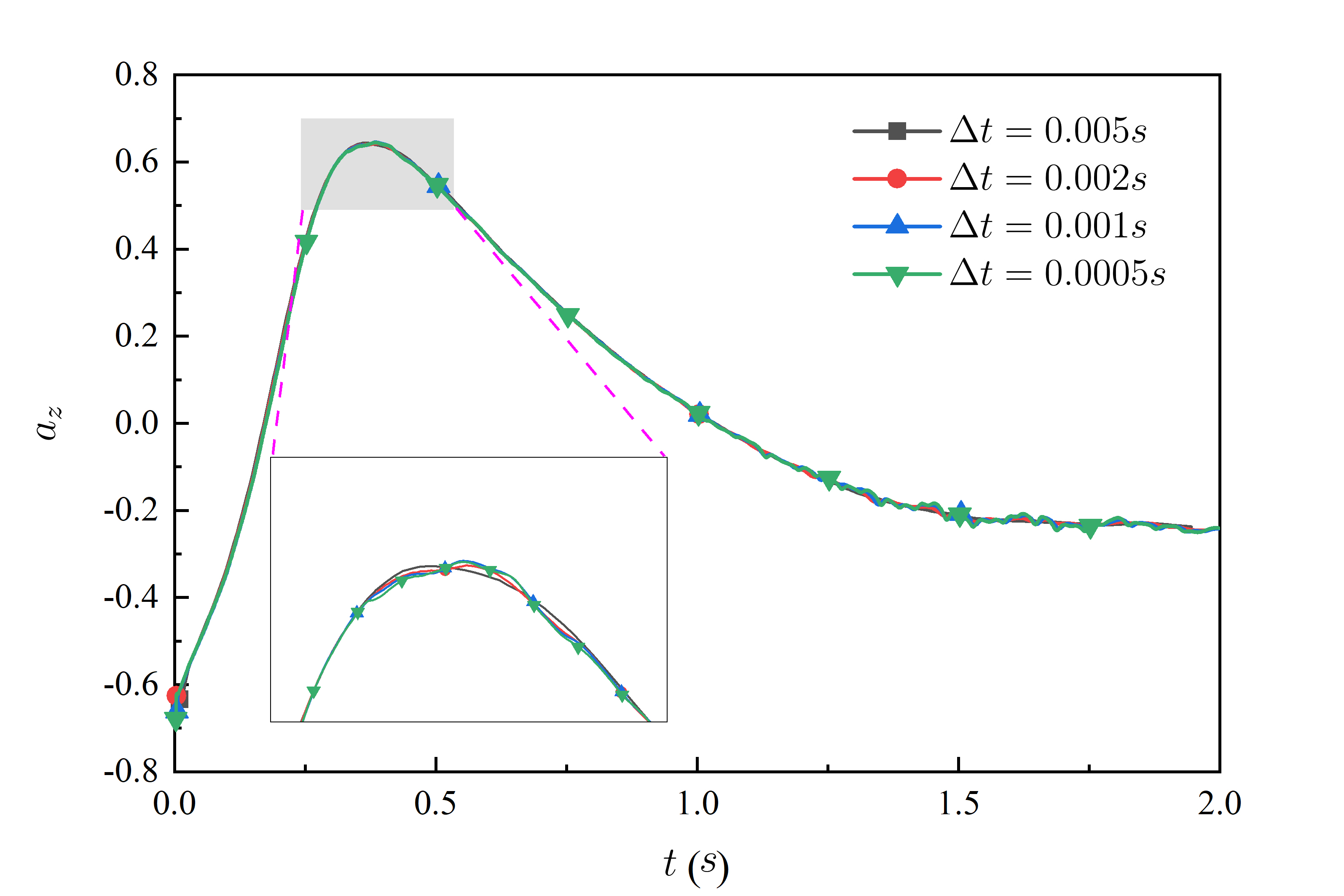}\\
\textbf{a)}\qquad\qquad\qquad\qquad\qquad\qquad\qquad\qquad\qquad\qquad\qquad\textbf{b)}
\caption{Grid and time-step size independence tests on dimensionless acceleration $a_z$: a) grid independence study; b) time-step independence study.}
\label{fig:Convergence}
\end{figure}

Another aim of the validation is to investigate the convergence of the solution when increasing grid resolution and refining time step size. To assess the effects of spatial discretization, a mesh refinement study for the case of amphibious aircraft with hydrofoil landing on still water has been conducted on three mesh systems with different resolution, as seen in Table~\ref{tab:Grid} in which $\Delta s$ represents the size of mesh on the surface of hydrofoil. Fig.~\ref{fig:Convergence}a shows the time histories of non-dimensional acceleration $a_z$ computed by the three meshes for the case of landing with $\upsilon_{x0}=48$ m/s and $\upsilon_{z0}=2$ m/s. Numerical uncertainty due to discretization errors on the prediction of the maximum acceleration is considered and corresponding grid convergence index (GCI) method is derived \citep{celik2008procedure,islam2021assessment,wang2021cfd}. The relevant results are presented in Table~\ref{tab:GridUncertainty} and the relatively low uncertainty can be observed being below 5$\%$, although oscillatory convergence are exhibited ($\epsilon_{32}/\epsilon_{21}<0$). Therefore the medium-size grid is chosen throughout the paper as a trade-off between computational expense and accuracy.

\begin{table}[hbt!]
\caption{Mesh resolution for grid convergence study.}
\centering
\setlength{\tabcolsep}{10mm}{
\begin{tabular}{ccc}
\hline\hline
No.& Mesh& $\Delta s$, m\\
\hline
1& Fine&     0.01\\
2& Medium&   0.02\\
3& Coarse&   0.04\\
\hline\hline
\end{tabular}}
\label{tab:Grid}
\end{table}

\begin{table}[hbt!]
\caption{Grid uncertainty estimation for water landing based on the maximum acceleration.}
\centering
\begin{tabular}{ccc}
\hline\hline
Property& & $a_{z\mathrm{max}}$\\
\hline
Output values& $\phi_1$ (fine)& 0.6368\\
& $\phi_2$ (medium)& 0.6428\\
& $\phi_3$ (coarse)& 0.6109\\
Refinement ratio& $r_{21}$& 2\\
& $r_{32}$& 2\\
Difference of estimation& $\epsilon_{21}$& 0.0061\\
& $\epsilon_{32}$& -0.0329\\
Order of accuracy& $p$& 2.4033\\
Extrapolated value& $\phi_\mathrm{ext}^{21}$& 0.6354\\
& $\phi_\mathrm{ext}^{32}$& 0.6502\\
Approximate relative error& $e_\mathrm{a}^{21}$& 0.0095\\
& $e_\mathrm{a}^{32}$& 0.0496\\
Extrapolated relative error& $e_\mathrm{ext}^{21}$& 0.0022\\
& $e_\mathrm{ext}^{32}$& 0.0114\\
Grid convergence index (GCI)& $GCI^{21}$& 0.0027\\
& $GCI^{32}$& 0.0144\\
Grid convergence index (GCI)& $GCI^{21}$& 0.0027\\
& $GCI^{32}$& 0.0144\\
Uncertainty& $U_1$& 0.27$\%$\\
& $U_2$& 1.44$\%$\\
Corrected uncertainty& $U_{1\mathrm{c}}$& 0.05$\%$\\
& $U_{2\mathrm{c}}$& 0.29$\%$\\
\hline\hline
\end{tabular}
\label{tab:GridUncertainty}
\end{table}

The time-step size refinement study is performed with four time-step sizes on the medium mesh for the same impacting condition. Fig.~\ref{fig:Convergence}b shows the time-varying acceleration $a_z$. It can be seen that \textcolor{black}{nearly the same} variation trends of time-varying acceleration are captured by all the computations with these four time-step sizes. Considering both the efficiency and the accuracy of the simulation, $\Delta t = 0.002$s is selected as a reasonable alternative for the following cases.

\section{Results and Discussion}
\label{sec:results}

\subsection{Analysis of Hydrofoil Geometric Parameters}
In order to evaluate the effect of hydrofoil geometric parameter on load reduction performance and obtain a hydrodynamically acceptable configuration for the following study on wave landing, only the pure fuselage with hydrofoil is first taken into consideration herein excluding struts, wing and tail wing components. Note that only two degrees of freedom (horizontal and vertical displacement) are allowed herein with a fixed pitching angle 5$^\circ$. The initial horizontal velocity and vertical velocity are 48m/s and 2m/s, respectively. For identification purposes, the original configuration and the configuration with hydrofoil are designated as OC and HC. Moreover, the rate of load reduction $\mu$ is defined as:

\begin{linenomath}
\begin{equation}
\mu = \dfrac{a_\mathrm{OC} - a_\mathrm{HC}}{a_\mathrm{OC}} \times 100\%
\end{equation}
\end{linenomath}

During water landing of an amphibious aircraft, the original configuration (OC) will experience the fuselage impacting stage, whereas the configuration with hydrofoil (HC) undergoes two stages: hydrofoil impacting and fuselage impacting. The hydrofoil geometric parameters significantly affect the hydrofoil and subsequent fuselage impact events. In the following investigations, the effect of area of hydrofoil is first discussed, followed by numerical analysis of the variations in sweep angle and dihedral angle of the hydrofoil.

\subsubsection{Effect of Static Load Coefficient}

\begin{figure}[hbt!]
\centering
\includegraphics[width=.49\textwidth]{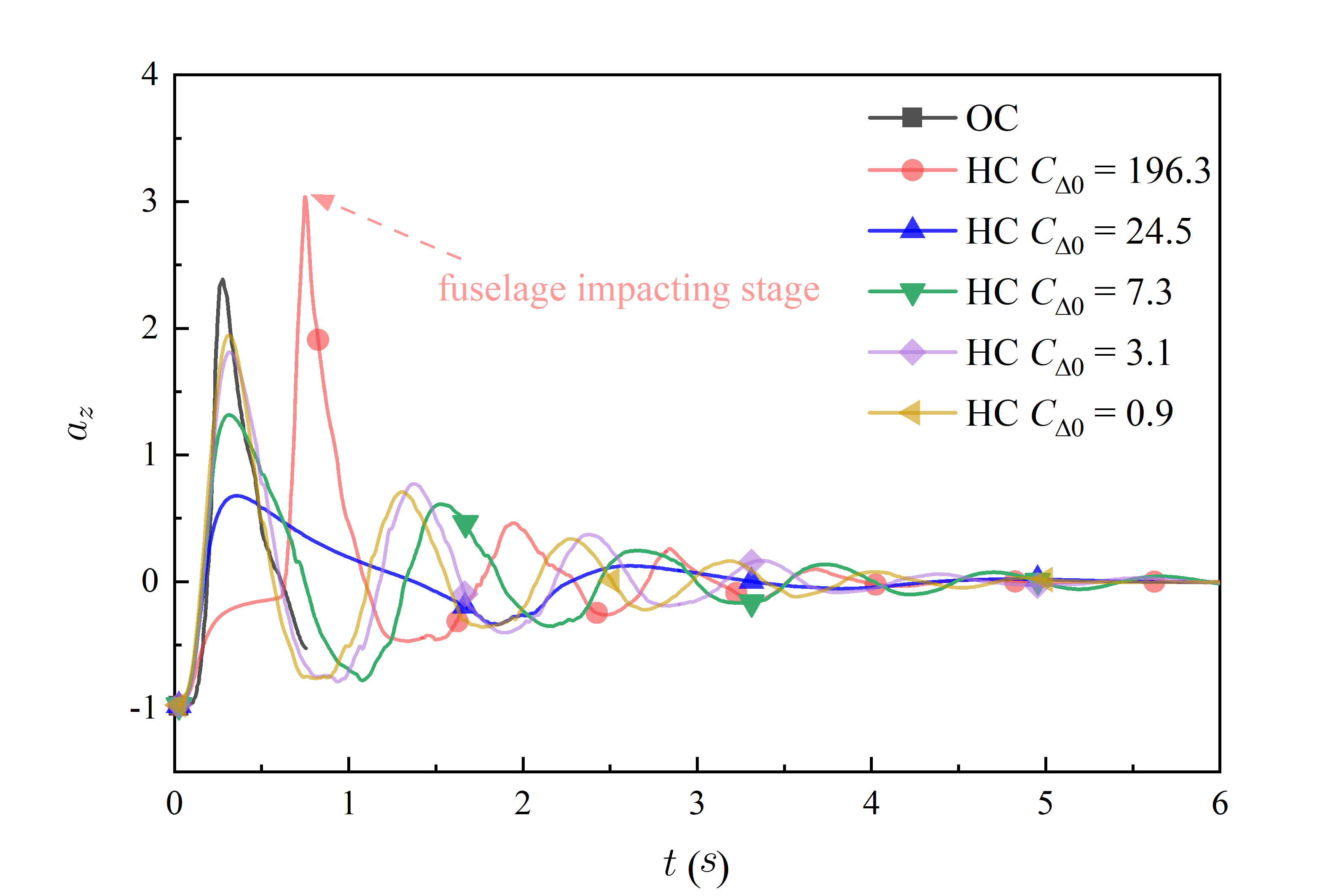}
\includegraphics[width=.49\textwidth]{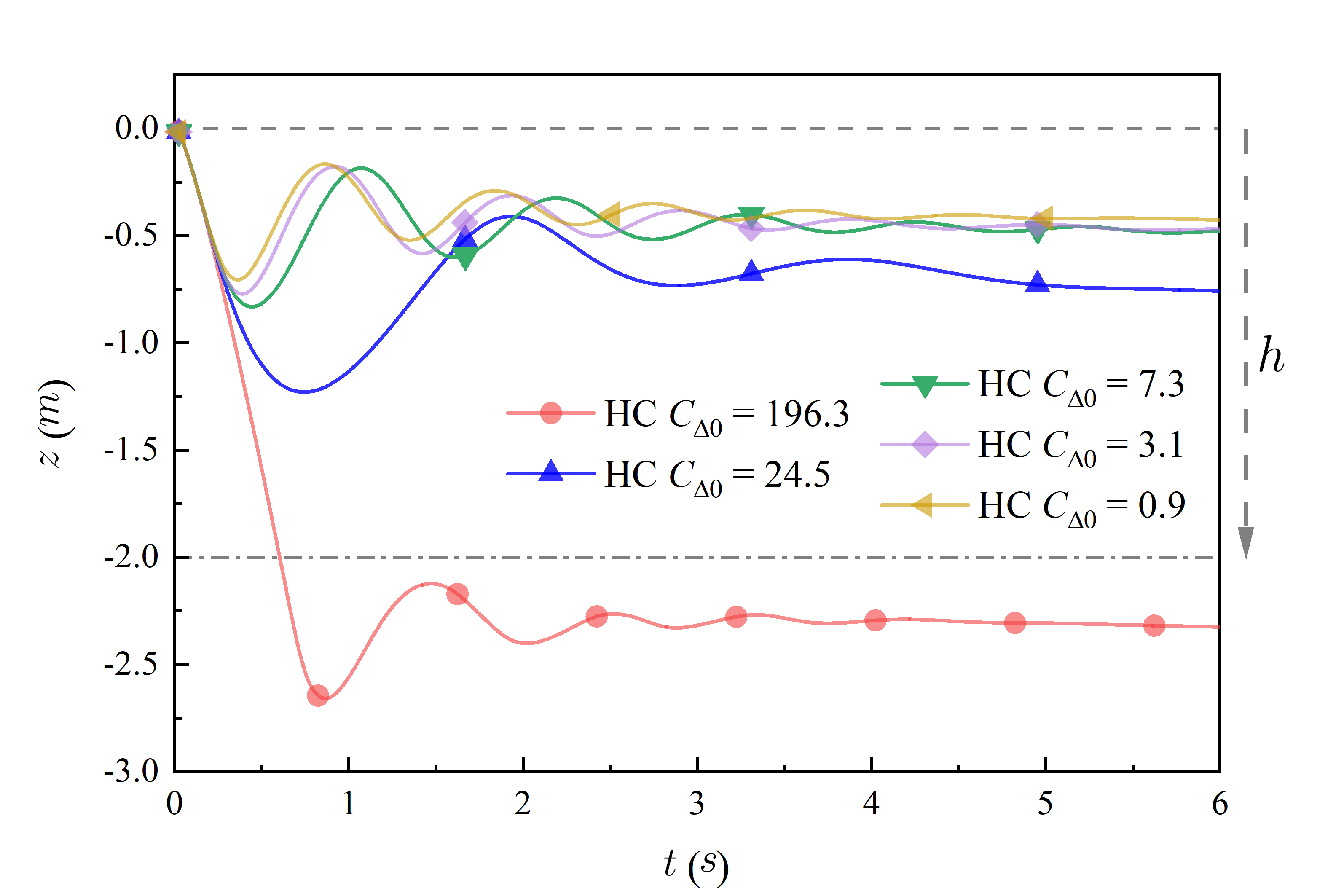}\\
\textbf{a)}\qquad\qquad\qquad\qquad\qquad\qquad\qquad\qquad\qquad\qquad\qquad\textbf{b)}
\caption{Time histories of dimensionless vertical acceleration $a_z$ and vertical displacement $z$ for various static load coefficients $C_{\Delta0}$: a) $a_z$; b) $z$.}
\label{fig:EffectAR}
\end{figure}

\begin{figure}[hbt!]
\centering
\includegraphics[width= 0.9\textwidth]{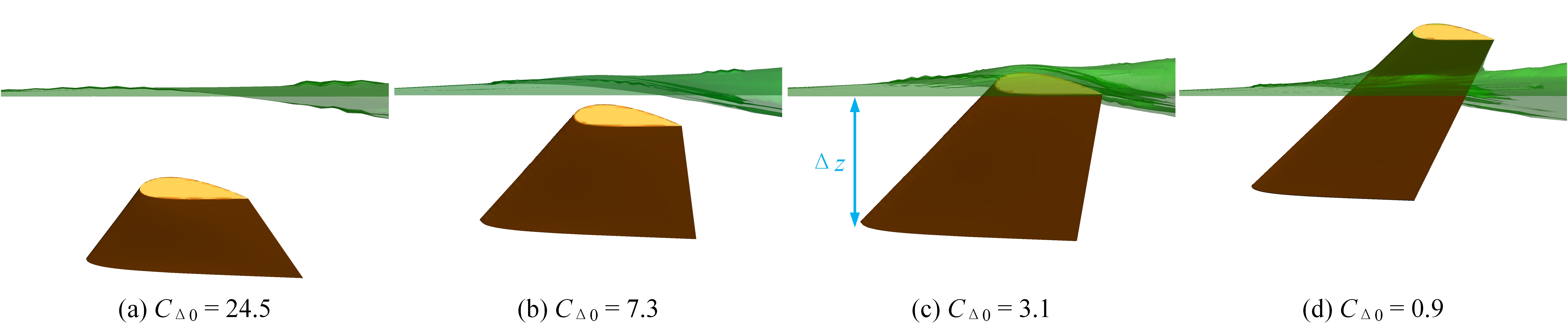}
\caption{Position of hydrofoil relative to the still water level for case HC $C_{\Delta0}=24.5$, 7.3, 3.1 and 0.9 when $a_z$ reaches maximum value.}
\label{fig:DifARAzmaxPosition}
\end{figure}

Time histories of dimensionless vertical acceleration in $z-$direction $a_z$ and vertical displacement for different static load coefficient $C_{\Delta0}$ are shown in Fig.~\ref{fig:EffectAR}. Note that both of sweep angle and dihedral angle of the hydrofoil are set to 20$^\circ$ in this study. At the first stage, which is the hydrofoil impacting stage, the value of $a_z$ grows immediately due to the sudden hydrodynamic impact forces exerted on the hydrofoil. The maximum value of $a_z$ approaches the result from OC case with decreasing $C_{\Delta0}$. During this period, the hydrofoil continuously moves downward as drawn in Fig.~\ref{fig:EffectAR}. It is worth noting that for the case $C_{\Delta0}=196.3$, the vertical displacement exceeds the height of struts and the fuselage starts to touch the water surface, causing a further increase in $a_z$ (see Fig.~\ref{fig:EffectAR}a). For cases where $C_{\Delta0}$ varies from 24.5 to 0.9, Fig.~\ref{fig:DifARAzmaxPosition} presents the relative positions between the hydrofoil and still water level when the maximal acceleration is achieved. It can be observed that the penetration depth $\Delta z$ decreases for smaller $C_{\Delta0}$ and the hydrofoil is not completely submerged when $C_{\Delta0}=0.9$.

\begin{figure}[hbt!]
\centering
\includegraphics[width= 0.9\textwidth]{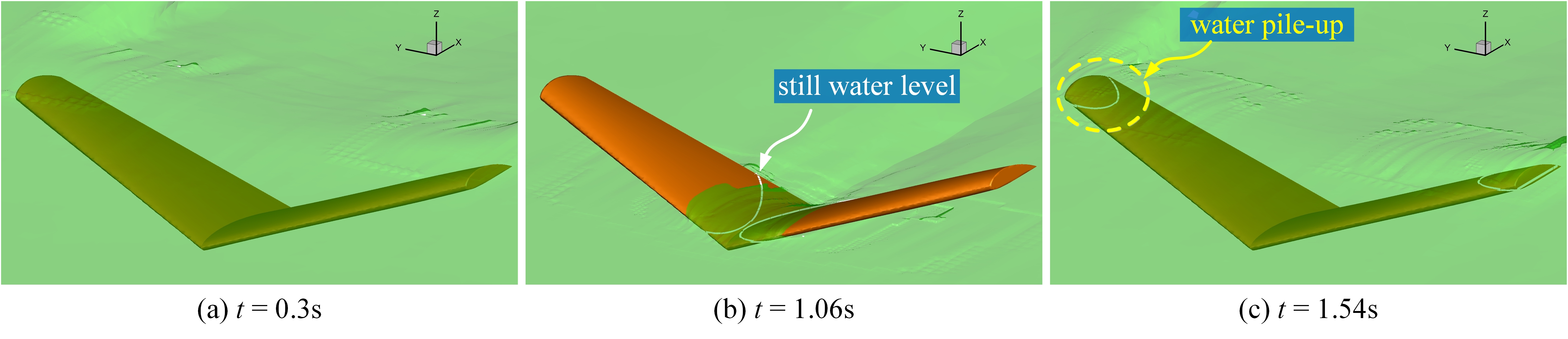}
\caption{Three time instants for case HC $C_{\Delta0}=24.5$ when $a_z$ experiences extreme values.}
\label{fig:AR1p5Motion}
\end{figure}

Several oscillations can be observed in $a_z$ and $z$ after the hydrofoil impacting stage due to the dynamic balance between additional hydrodynamic force generated by the hydrofoil and the gravity of aircraft. Taking $C_{\Delta0}=24.5$ as an example case, three moments when values of $a_z$ obtain the first peak, the first valley and the second peak are plotted in Fig.~\ref{fig:AR1p5Motion}, where the alternate behaviour of submergence and emergence of hydrofoil relative to the free surface can be observed graphically. By looking at the dynamic behaviour of the case $C_{\Delta0}=24.5$ (marked as blue line in Fig.~\ref{fig:EffectAR}), $a_z$ is always below 1$g$ with a high rate $\mu=60\%$ compared with the OC case, together with a moderate change in $z$. Therefore, it can be concluded that a static load coefficient $C_{\Delta0}$ around 24.5 is more suitable for the investigated range $[0.9, 196.3]$ and is selected for the following discussion.

\subsubsection{Effect of Sweep Angle and Dihedral Angle}

\begin{figure}[hbt!]
\centering
\includegraphics[width=0.49 \textwidth]{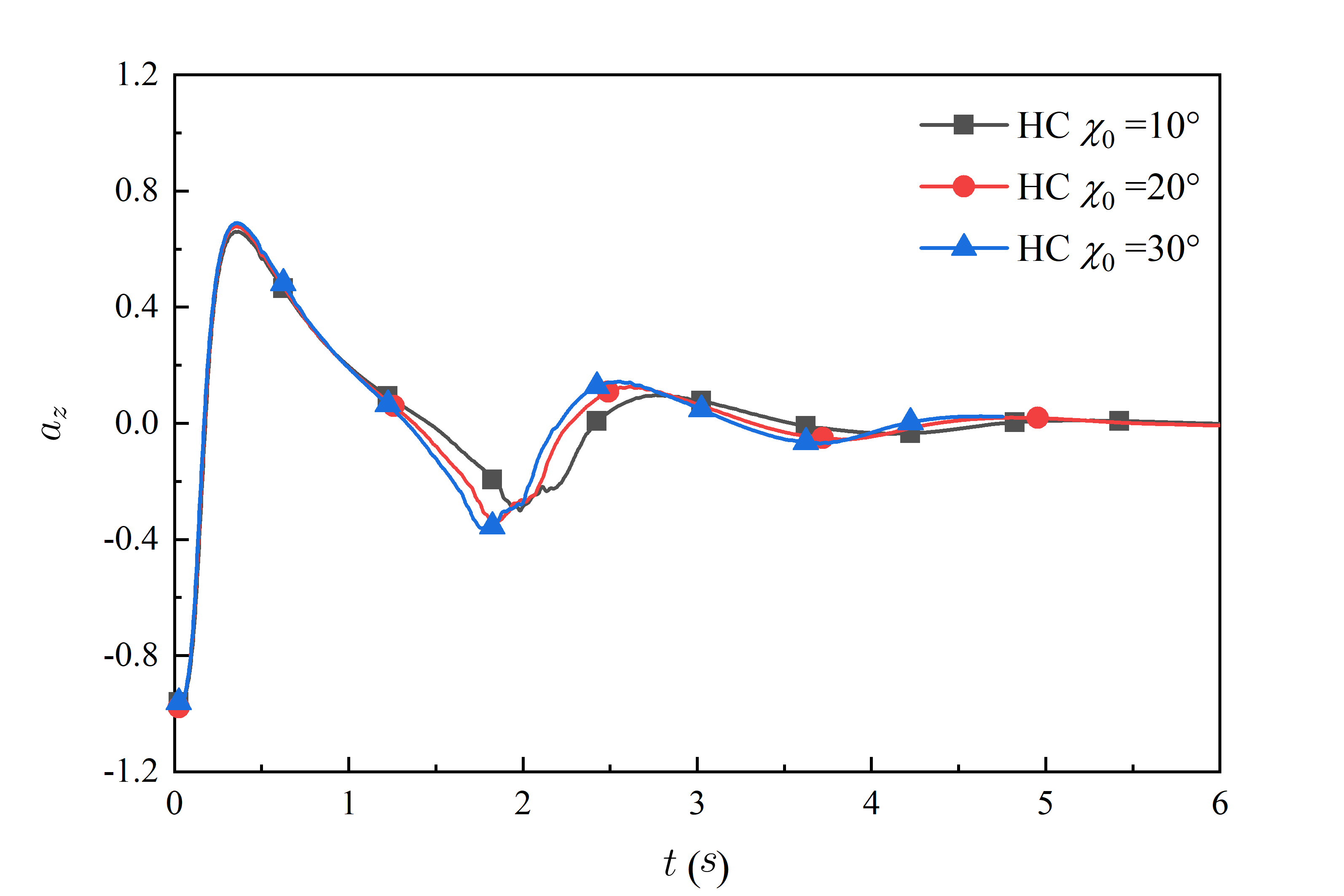}
\caption{Time histories of dimensionless vertical acceleration for different sweep angles $\chi_0$.}
\label{fig:EffectSA}
\end{figure}

Subsequently, three different cases of sweep angle $\chi_0$ (namely $10^\circ, 20^\circ, 30^\circ$) are investigated with a static load coefficient $C_{\Delta0}=24.5$ and a dihedral angle $\psi = 20^\circ$. Time histories of the dimensionless vertical acceleration $a_z$ are shown in Fig.~\ref{fig:EffectSA}. The results indicate a marginal effect of the sweep angle on the vertical acceleration for the investigated static load coefficient and dihedral angle, as similar trends are observed for all three cases.

\begin{figure}[hbt!]
\centering
\includegraphics[width=.49\textwidth]{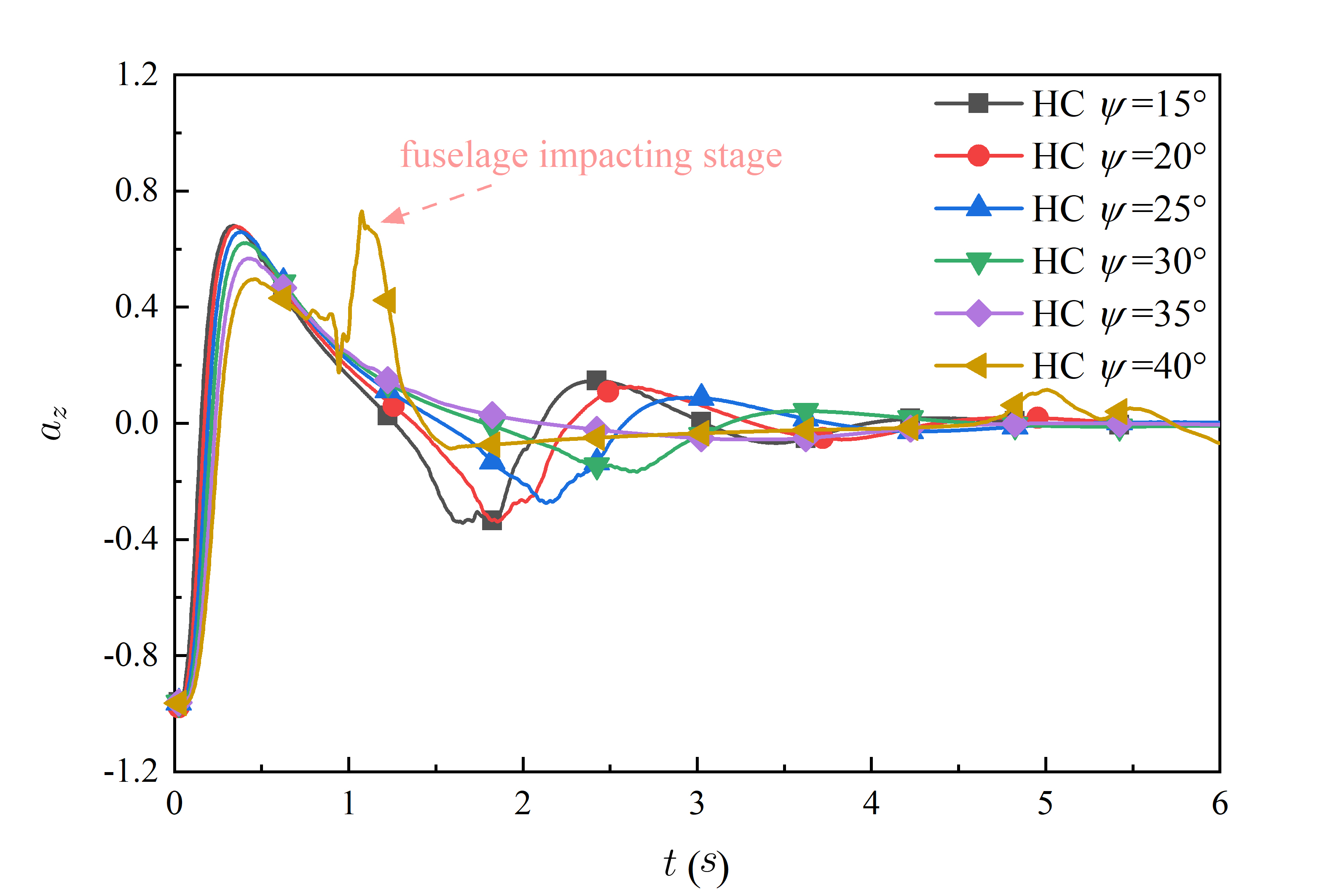}
\includegraphics[width=.49\textwidth]{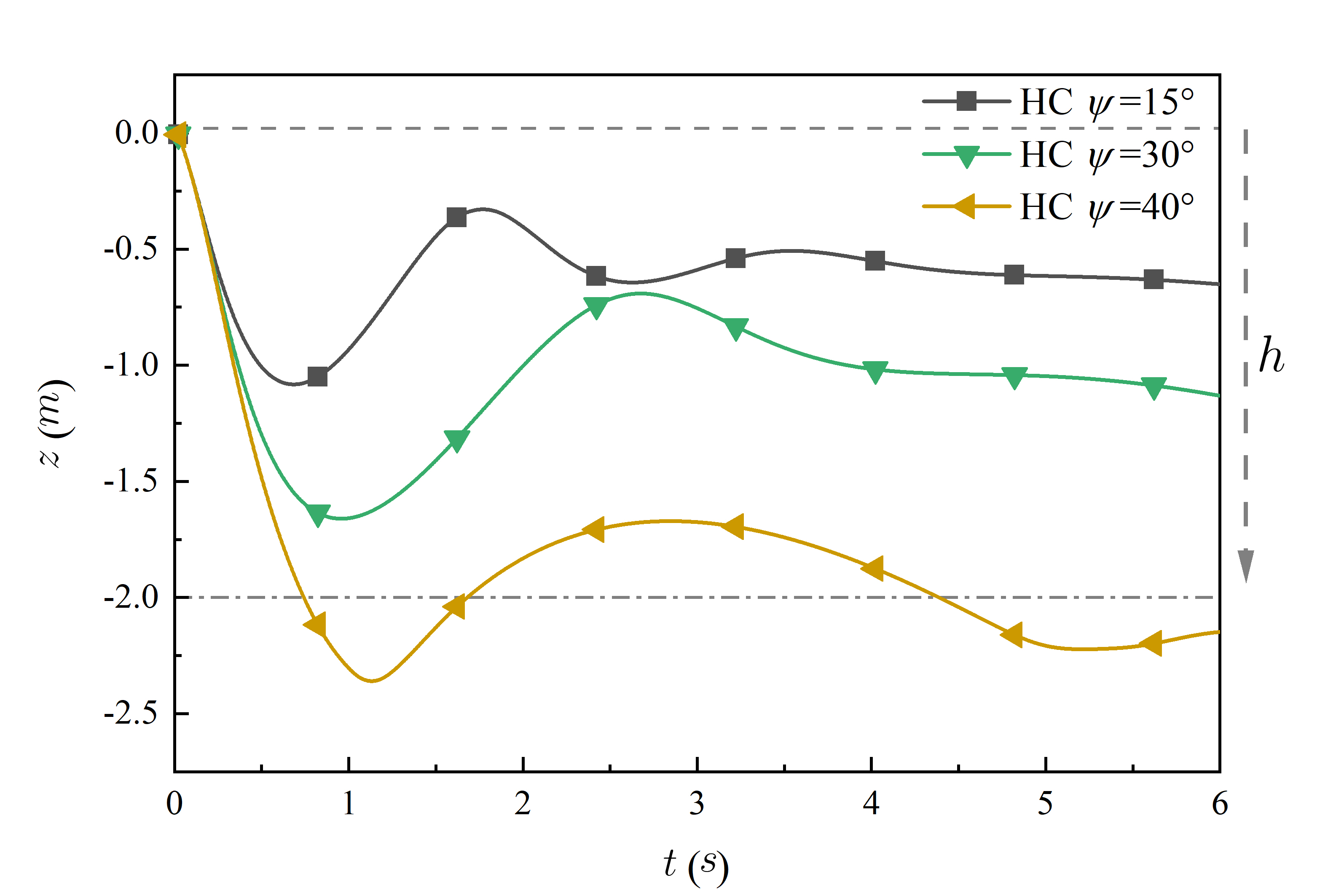}\\
\textbf{a)}\qquad\qquad\qquad\qquad\qquad\qquad\qquad\qquad\qquad\qquad\qquad\textbf{b)}
\caption{Time histories of dimensionless vertical acceleration $a_z$ and vertical displacement $z$ for different dihedral angles $\psi$: a) $a_z$; b) $z$.}
\label{fig:EffectDA}
\end{figure}

Regarding the investigation of the effect of dihedral angle $\psi$, several simulations have been conducted with a constant static load coefficient $C_{\Delta0}=24.5$, sweep angle $\chi_0 =20^\circ$ and varying dihedral angle $\psi$ from 15$^\circ$ to 40$^\circ$. Time histories of dimensionless acceleration in $z-$direction $a_z$ and vertical displacement $z$ are depicted in Fig.~\ref{fig:EffectDA}. The data indicate that an increase in $\psi$ causes a minor reduction in the maximal value of $a_z$, which is consistent with the law of deadrise angle on wedge-shaped section water impact \citep{zhao1993water,iafrati2000hydroelastic}, and exhibits a further penetration depth of the body. The height of struts $h$ in the present study is setup as 2m, and the maximum change of $z$ in case $\psi=40^\circ$ beyond 2m (see Fig.~\ref{fig:EffectDA}b), leading to the fuselage hitting the free surface and resulting in a sharp increase in $a_z$, as shown in Fig.~\ref{fig:EffectDA}a. It can be concluded that there is a limited range of dihedral angle $\psi$ for the specific $C_{\Delta0}$.
Based on the above investigations, a set of compromised configuration parameter of the hydrofoil with a static load coefficient $C_{\Delta0}=24.5$, $\chi_0 =20^\circ$, and dihedral angle $\psi =35^\circ$ is chosen for further study of load reduction performance of an amphibious aircraft landing on still/wavy water.

\subsection{Load Reduction for the Whole Amphibious Aircraft}
To further study the load reduction performance of the amphibious aircraft using hydrofoil, several generic conditions that the aircraft could experience are numerically investigated, such as different descending velocity for landing on still water, the effect of wave length and wave steepness when encountering wavy water.

\subsubsection{Effect of Descent Velocity on Still Water}
\begin{figure}[hbt!]
\centering
\includegraphics[width=.49\textwidth]{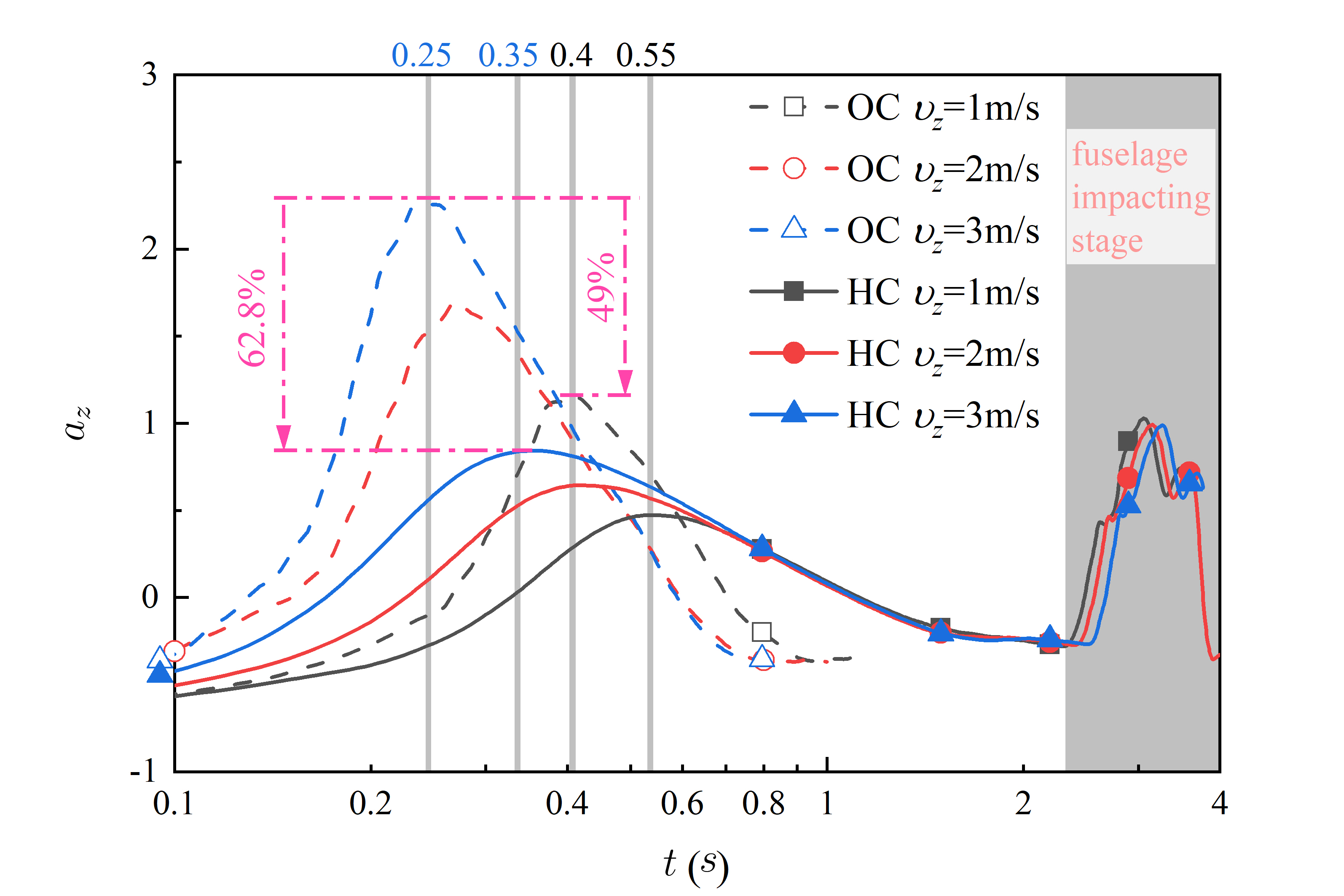}
\includegraphics[width=.49\textwidth]{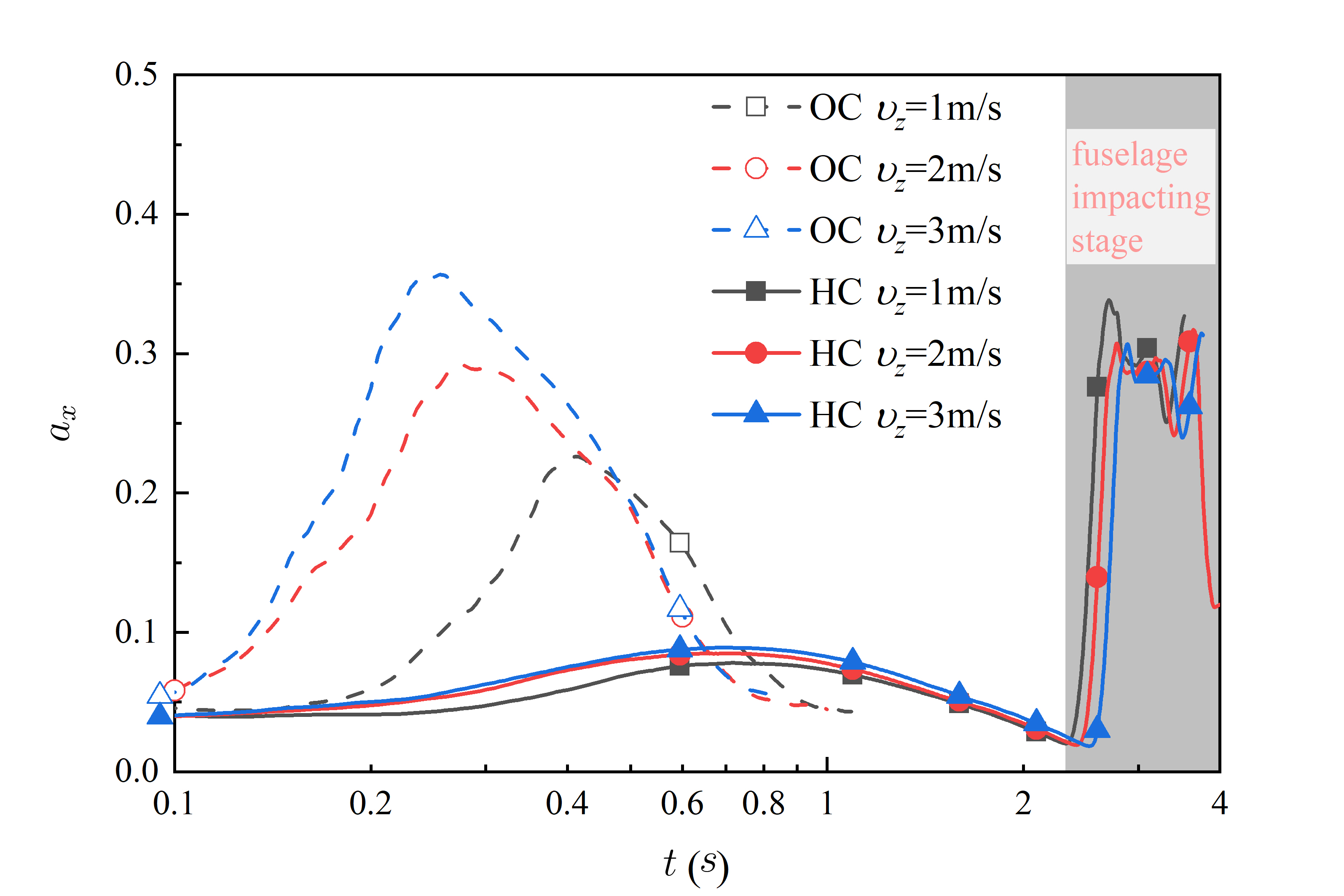}\\
\textbf{a)}\qquad\qquad\qquad\qquad\qquad\qquad\qquad\qquad\qquad\qquad\qquad\textbf{b)}
\caption{Time histories of acceleration for original and hydrofoil configurations (OC and HC) with various initial descent velocities $\upsilon_{z0}$: a) $z-$direction; b) $x-$direction.}
\label{fig:StillCompAcc}
\end{figure}

\begin{figure}[hbt!]
\centering
\includegraphics[width=0.49 \textwidth]{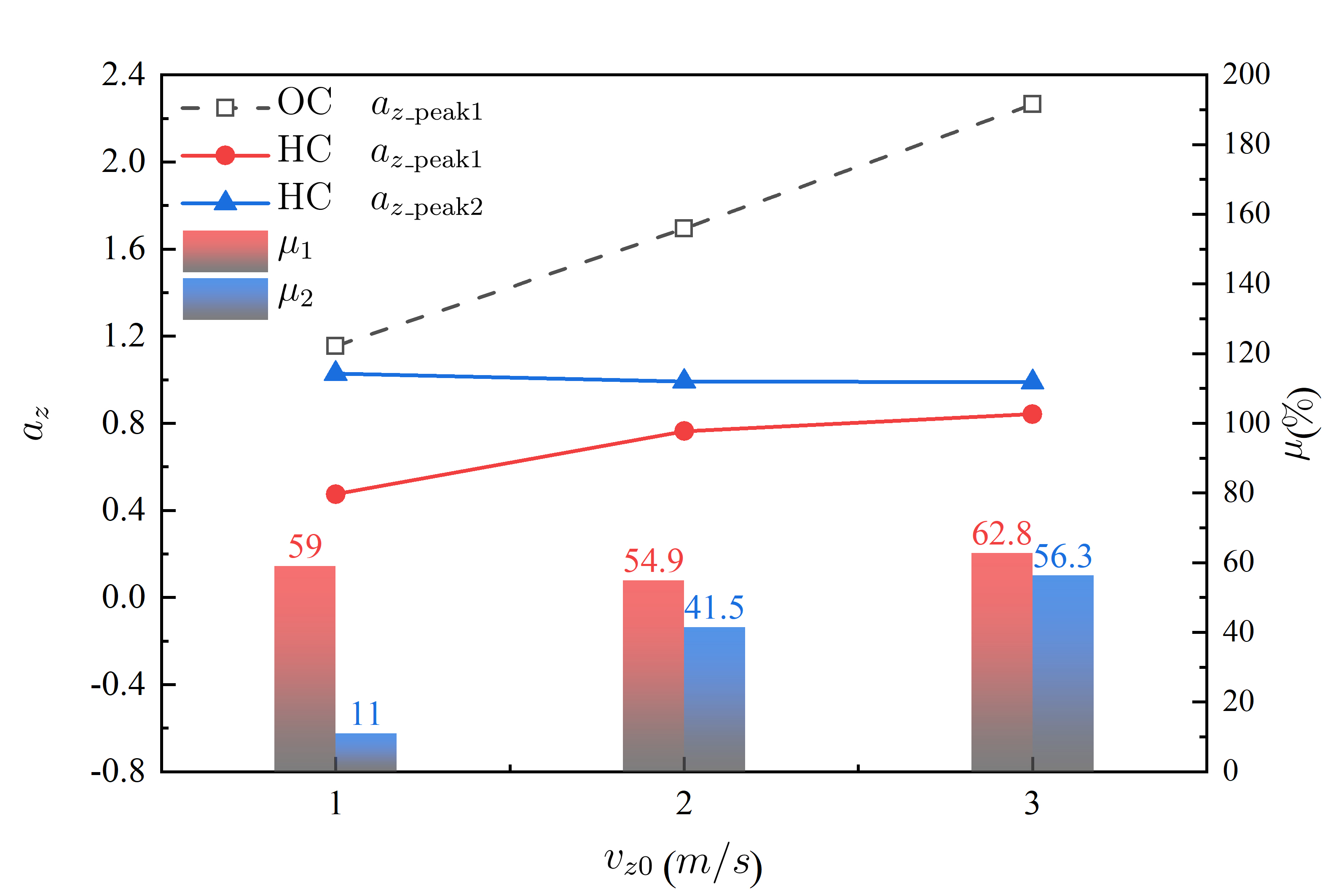}
\caption{Comparison of peak accelerations in $z$-direction for OC and HC cases with various initial descent velocities $\upsilon_{z0}$ landing on still water.}
\label{fig:StillCompPeak}
\end{figure}

Three different initial descent velocities $\upsilon_{z0}$ are discussed herein, i.e., 1m/s, 2m/s and 3m/s. The results of acceleration in $z-$ and $x-$direction are presented in Fig.~\ref{fig:StillCompAcc} and compared with the data from the original configuration (OC). Fig.~\ref{fig:StillCompPeak} shows a comparison of the vertical peak acceleration associated with corresponding load reducing rate $\mu$ for both the original configuration (OC) and the configuration equipped with hydrofoil (HC). Note that the initial horizontal velocity $\upsilon_{x0}$ and pitching angle $\theta$ are 48 m/s and 5$^\circ$, respectively. Looking at the results of the original configuration, both of $a_z$ and $a_x$ reduce with decreasing $\upsilon_{z0}$. Similar phenomenon can be found for the HC cases varying $\upsilon_{z0}$ at the hydrofoil impacting stage and all values are below $1g$. It is worth noting that the result from the case of HC with $\upsilon_{z0}=3$ m/s is even less than that from the OC case with $\upsilon_{z0}=1$ m/s, indicating that the significant contribution of hydrofoil to the reduction in accelerations, which is more effective than varying initial velocity. Furthermore, for the cases with hydrofoil, accelerations exhibit a second peak period (see Fig.~\ref{fig:StillCompAcc}a and \ref{fig:StillCompAcc}b) due to the fuselage impacting stage. The load reduction rate $\mu_1$ is greater than $55\%$ at the first impacting phase. However, in the same impacting stage, that is fuselage impacting stage, $\mu_2$ exhibits a decreasing trend from $56.3\%$ to $11.0\%$ as $\upsilon_{z0}$ reduces.

\begin{figure}[hbt!]
\centering
\includegraphics[width=0.9\textwidth]{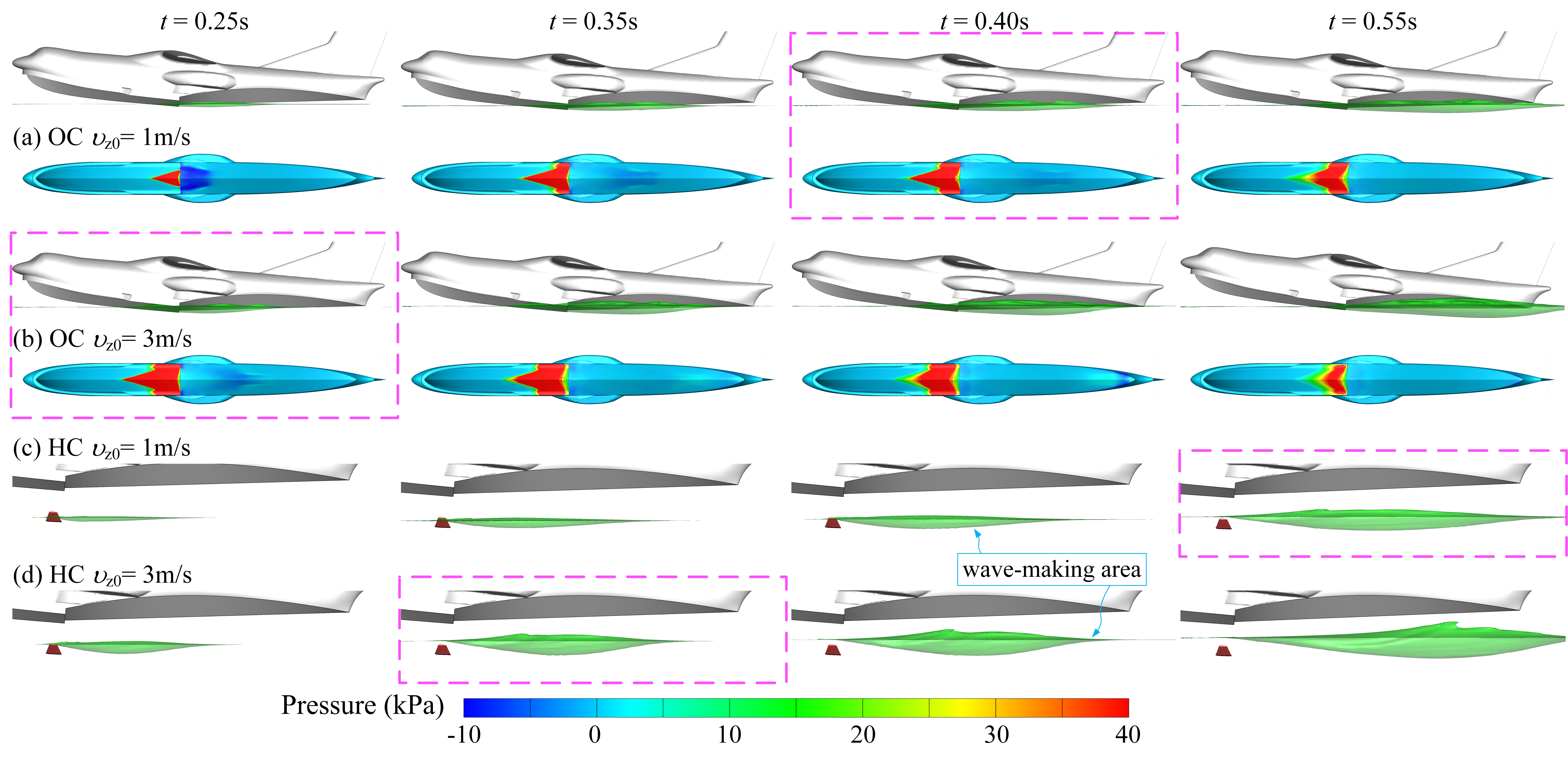}
\caption{Landing motions and pressure distribution at the bottom of the fuselage for OC and HC cases at $\upsilon_{z0}$=1m/s and 3m/s.}
\label{fig:ContourDifVz}
\end{figure}

\begin{figure}[hbt!]
\centering
\includegraphics[width=0.56\textwidth]{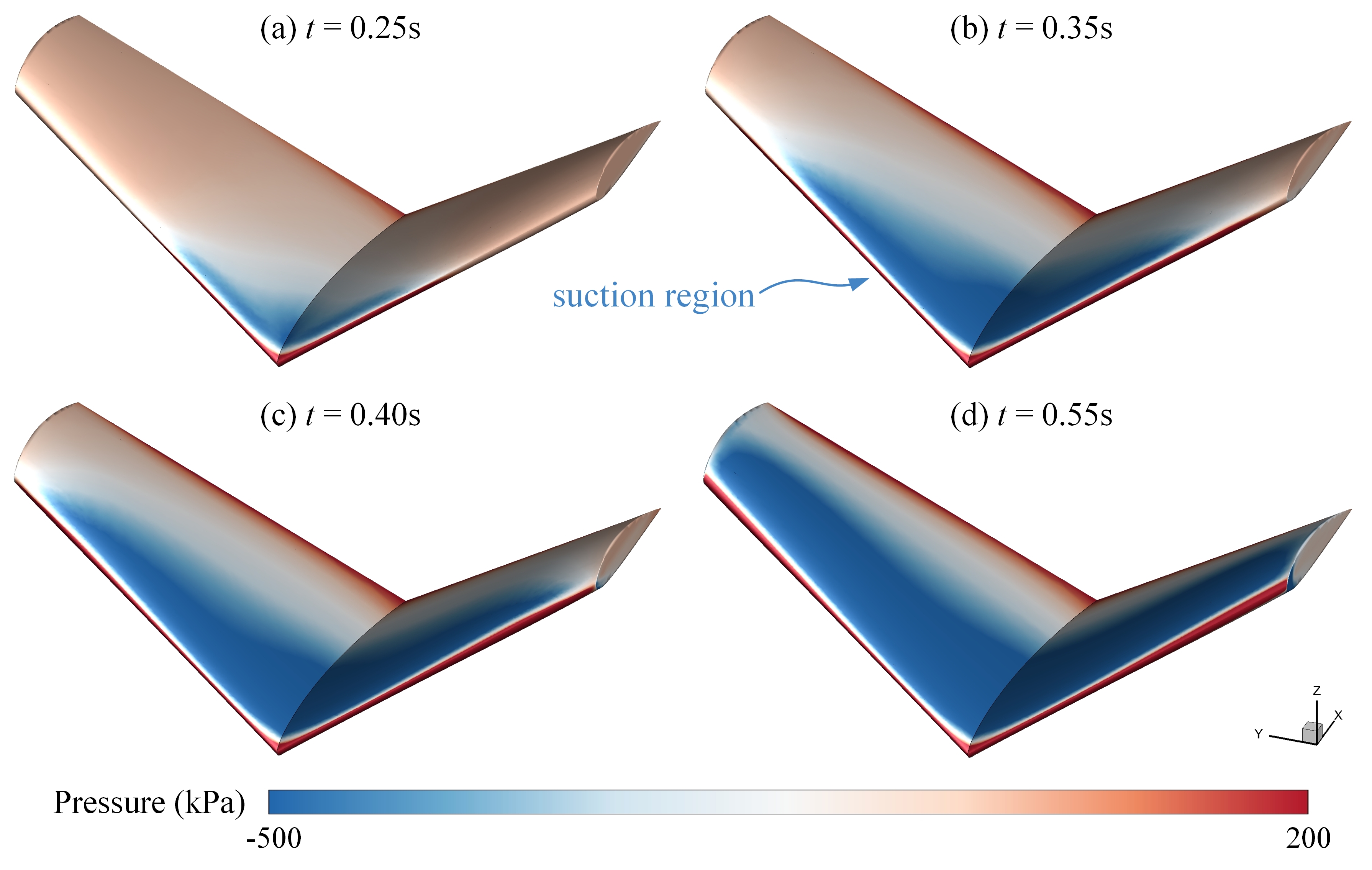}
\caption{Four snapshots of pressure distribution on the hydrofoil for the case HC $\upsilon_{z0}$=1m/s.}
\label{fig:HCvz1FoilPressure}
\end{figure}

Fig.~\ref{fig:ContourDifVz} illustrates the water surface deformation around the body and pressure distribution at the bottom of the aircraft for both OC and HC cases at $\upsilon_{z0}=1$m/s and 3m/s. Note that the moment when $a_z$ reaches its first peak value is marked with the magenta dashed rectangle. For the case of OC, the main fuselage portion striking with the water surface is the region over the forebody near the step, resulting in a triangle-shaped region of positive pressure. The high-pressure region becomes smaller in size and reduces in magnitude when decreasing $\upsilon_{z0}$. Moving to HC cases, a wave-making area is formed with the penetration of hydrofoil. It can be found that the first maximal acceleration occurs when the hydrofoil completely submerges and moves away from the wave-making area, which is similar to the phenomenon observed in Fig.~\ref{fig:DifARAzmaxPosition}. As it can be seen in Fig.~\ref{fig:HCvz1FoilPressure}, the size of low-pressure area on the suction surface becomes larger with the penetration of hydrofoil indicating an gradual increase in the resultant hydrodynamic lift force, which can be used to explain the occurrence of maximal acceleration. \textcolor{black}{It is worth noting that as the cavitation model \citep{schnerr2001physical} is not considered in the present numerical simulations, bubble or subsequent ventilation phenomena cannot be observed on the upper surface of the hydrofoil.} Moreover, a higher initial vertical velocity leads to a shorter time period before the first acceleration peak is reached, as presented in Fig.~\ref{fig:StillCompAcc}a, since the hydrofoil has less time to escape from the wave-making area.

\subsubsection{Effect of Wave Length}
\begin{figure}[hbt!]
\centering
\includegraphics[width=.49\textwidth]{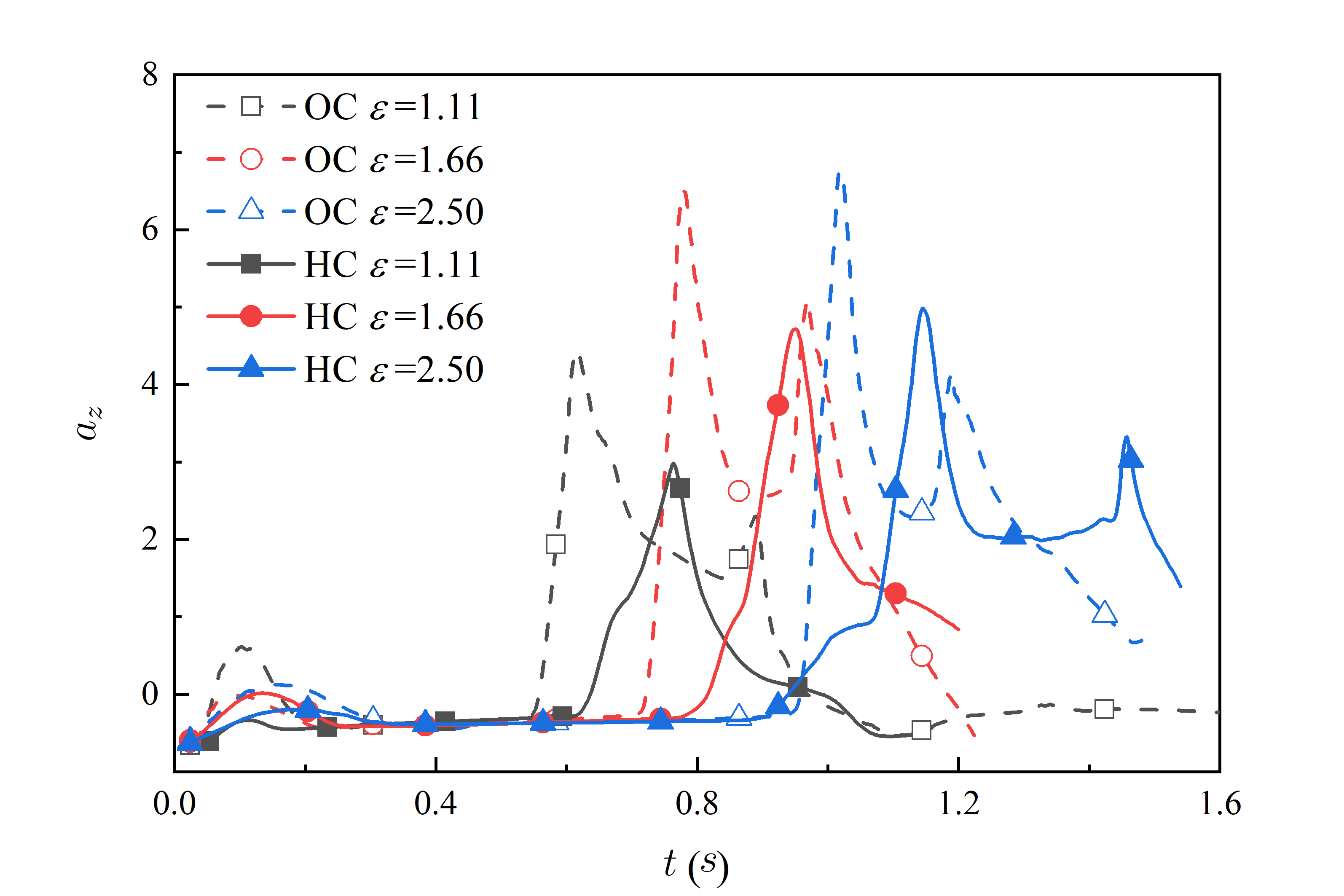}
\includegraphics[width=.49\textwidth]{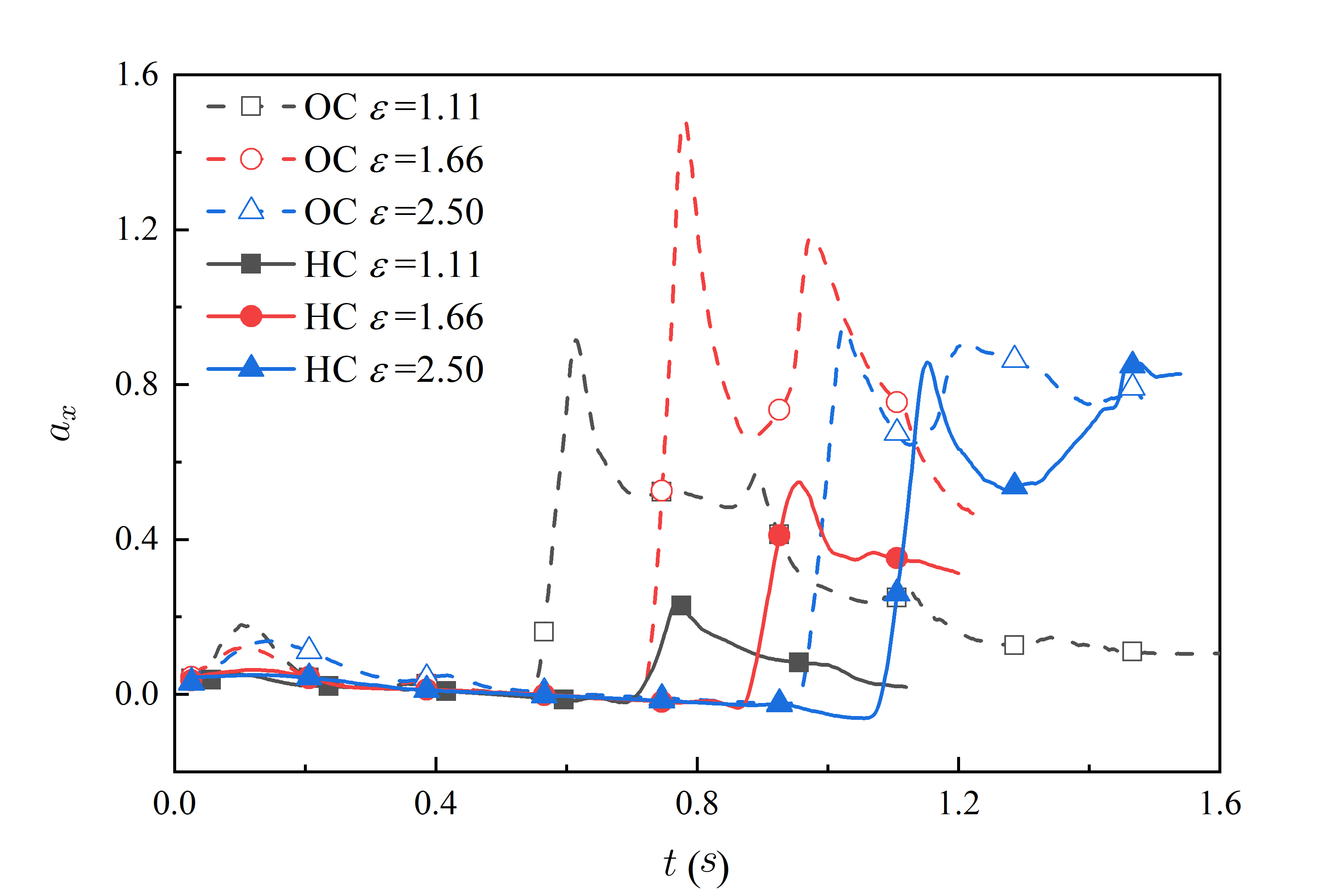}\\
\textbf{a)}\qquad\qquad\qquad\qquad\qquad\qquad\qquad\qquad\qquad\qquad\qquad\textbf{b)}
\caption{Time histories of acceleration for original and hydrofoil configurations (OC and HC) for different dimensionless wave lengths $\varepsilon$: a) $z-$direction; b) $x-$direction.}
\label{fig:WaveLengthCompAcc}
\end{figure}

\begin{figure}[hbt!]
\centering
\includegraphics[width=0.49 \textwidth]{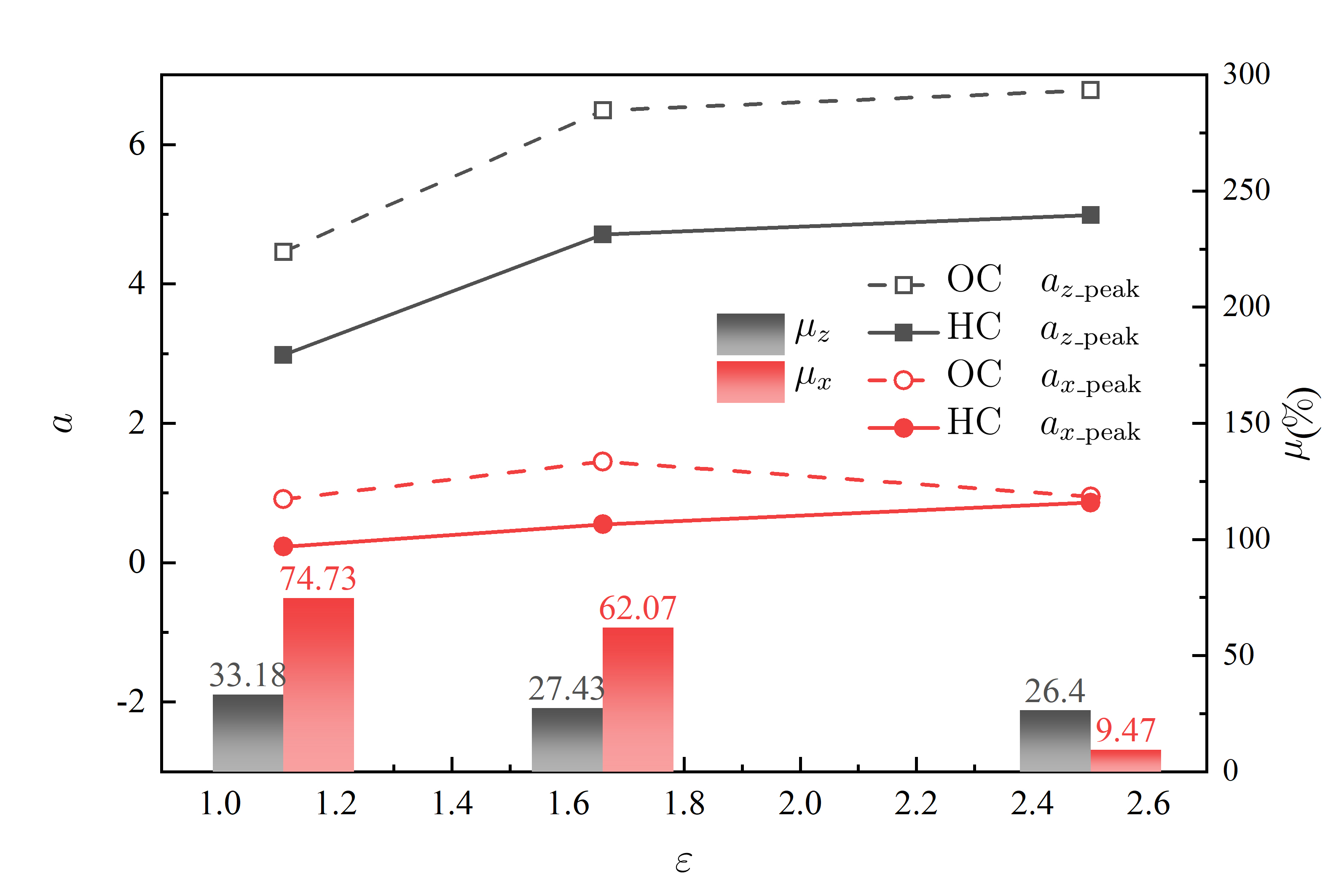}
\caption{Comparison of maximal accelerations in $z-$ and $x-$direction for OC and HC cases landing on crest at various dimensionless wave lengths $\varepsilon$.}
\label{fig:WaveLengthCompPeak}
\end{figure}

The possibility of landing on wavy water poses a significant challenge for the amphibious aircraft, making it necessary to investigate the hydrodynamic characteristics during such conditions and evaluate the effectiveness of load reduction with the help of hydrofoil. The effect of wave length is first explored with constant wave height $H$=3m. Note that crest is selected as the first contacting position. Considering the length of the fuselage $L$, three different dimensionless wave lengths $\varepsilon$ are considered, i.e., 1.11, 1.66 and 2.50, which is defined as $\varepsilon=\lambda_\mathrm{w}/L$. Fig.~\ref{fig:WaveLengthCompAcc} presents the time histories of vertical acceleration $a_z$ and horizontal acceleration $a_x$ for OC and HC under different wave lengths conditions. The results of maximal acceleration for various cases are also highlighted in Fig.~\ref{fig:WaveLengthCompPeak}. It can be observed that the load reduction rate $\mu_z$ in terms of $a_z$ decreases slightly from $33.18\%$ to $26.40\%$ as $\varepsilon$ increases, whereas $\mu_x$ drops dramatically below $10\%$.
These results demonstrate that the efficiency of the hydrofoil in reducing the load is more sensitive to changes in wave length for $a_x$ than for $a_z$. Overall, the importance of considering the effects of wave length on amphibious aircraft landing performance and the potential benefits of using hydrofoil to mitigate the impact of landing on wavy water has to be highlighted.

\begin{figure}[hbt!]
\centering
\includegraphics[width=0.9\textwidth]{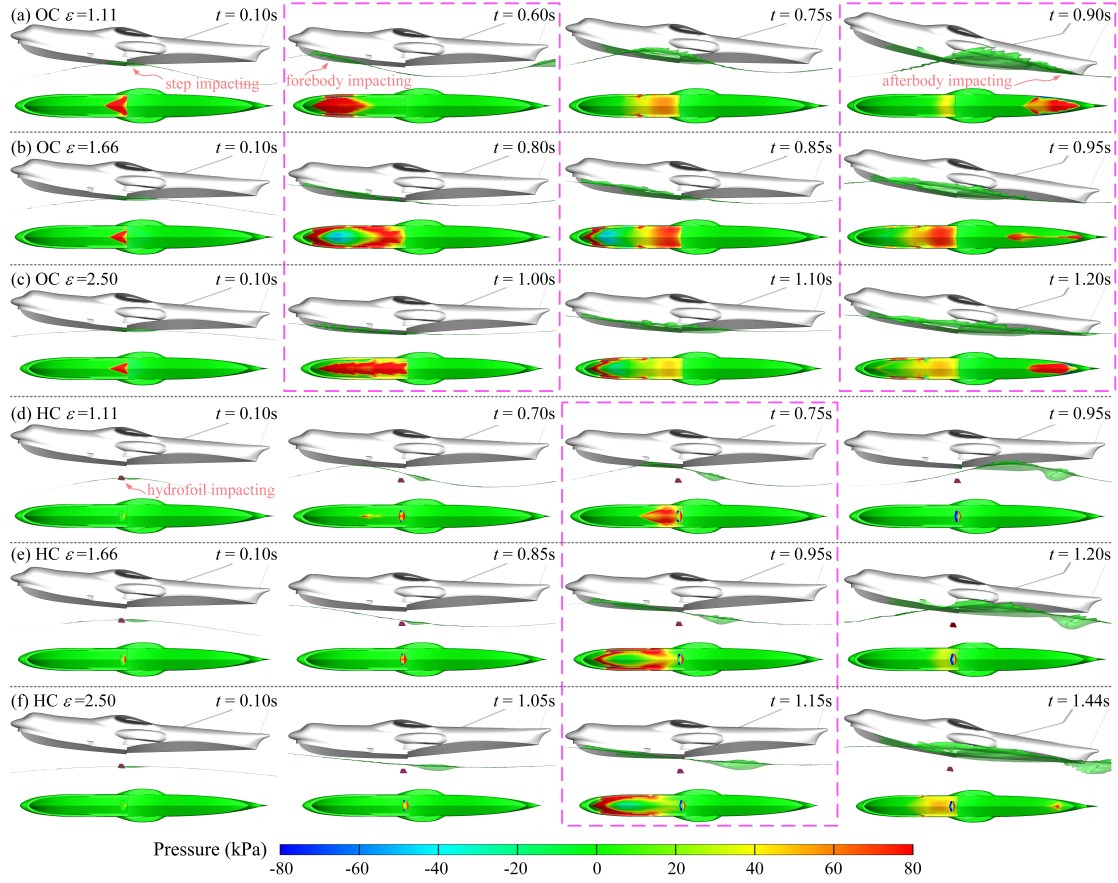}
\caption{Landing motions and pressure distribution around the peak vertical acceleration for various wave length $\varepsilon$.}
\label{fig:ContourDifWaveLength}
\end{figure}

\begin{figure}[hbt!]
\centering
\includegraphics[width=0.49 \textwidth]{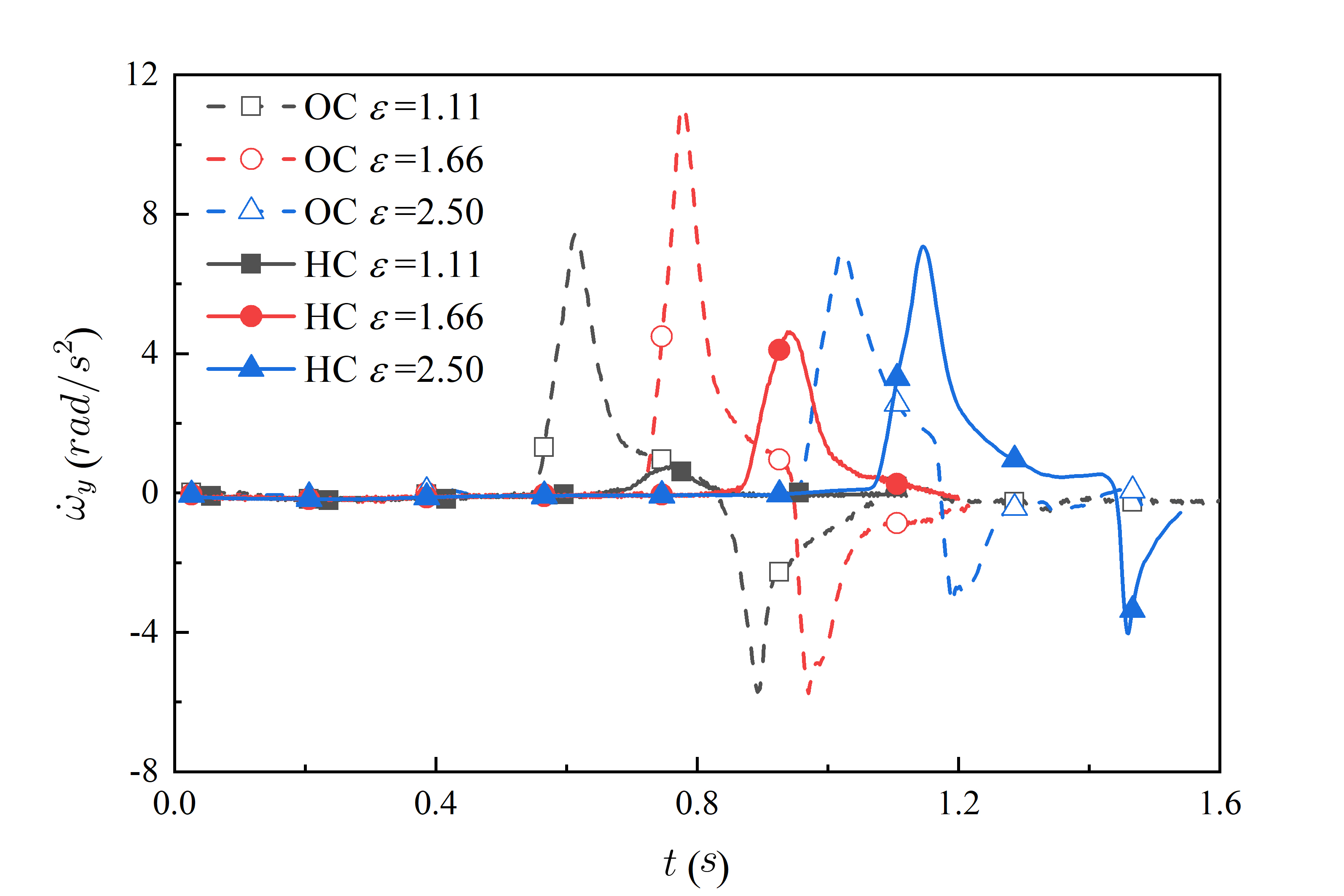}
\caption{Time histories of angular accelerations around $y-$direction for OC and HC cases landing on crest with different dimensionless wave lengths $\varepsilon$.}
\label{fig:WaveLengthCompOmega}
\end{figure}

In order to achieve a better comprehension of the effect of wave length on accelerations and corresponding load reduction performance, several time instants of the landing event for both HC and OC cases together with pressure distribution at the bottom of the aircraft are drawn in Fig.~\ref{fig:ContourDifWaveLength}, where moments when $a_z$ reaches its peak value are highlighted by the magenta dashed rectangle. At the onset of landing on crest for the OC case, the step hits the water somewhat and then flies clear of the first wave. During this period, only little hydrodynamic force is produced by the interaction between body and water, exhibiting slight fluctuations on both $a_z$ and $a_x$ (see Fig.~\ref{fig:WaveLengthCompAcc}). Generally, the maximum accelerations do not occur at the first contact. In the ensuing motion, the forebody of the aircraft encounters the rising position of the second oncoming wave with a high pressure zone appeared at the bottom, causing a sharp increase on both $a_z$ and $a_x$ until reaching the corresponding maximal values. Meanwhile, as the forebody is first impacted with rising wave, a pitching-up moment is generated as shown in Fig.~\ref{fig:WaveLengthCompOmega} where the angular accelerations around $y-$direction $\dot{\omega}_y$ for six cases are collected. Then, due to the combination of forward and pitching-up movement, the afterbody touches the water surface resulting a second peak on accelerations and a pitching-down moment (see Fig.~\ref{fig:WaveLengthCompAcc} and \ref{fig:WaveLengthCompOmega}). It can be observed that increasing the wave length contributes to a slight increase among $a_z$, $a_x$ and $\dot{\omega}_y$, whereas for OC $\varepsilon=2.50$ case, the results of $a_x$ and $\dot{\omega}_y$ exhibit a decreasing trend. Such difference can be schematically explained from Fig.~\ref{fig:ContourDifWaveLength} (a) $t=$0.60s, (b) $t=$0.80s and (c) $t=$1.00s where the forebody impacts the rising wave gradually close to the trough with smaller size of high pressure region moving to the step. A smaller relative velocity between free surface and fuselage is obtained since the elevation velocity of wave surface is approximately zero at trough \citep{zhang2023numerical}.

\begin{figure}[hbt!]
\centering
\includegraphics[width=0.49 \textwidth]{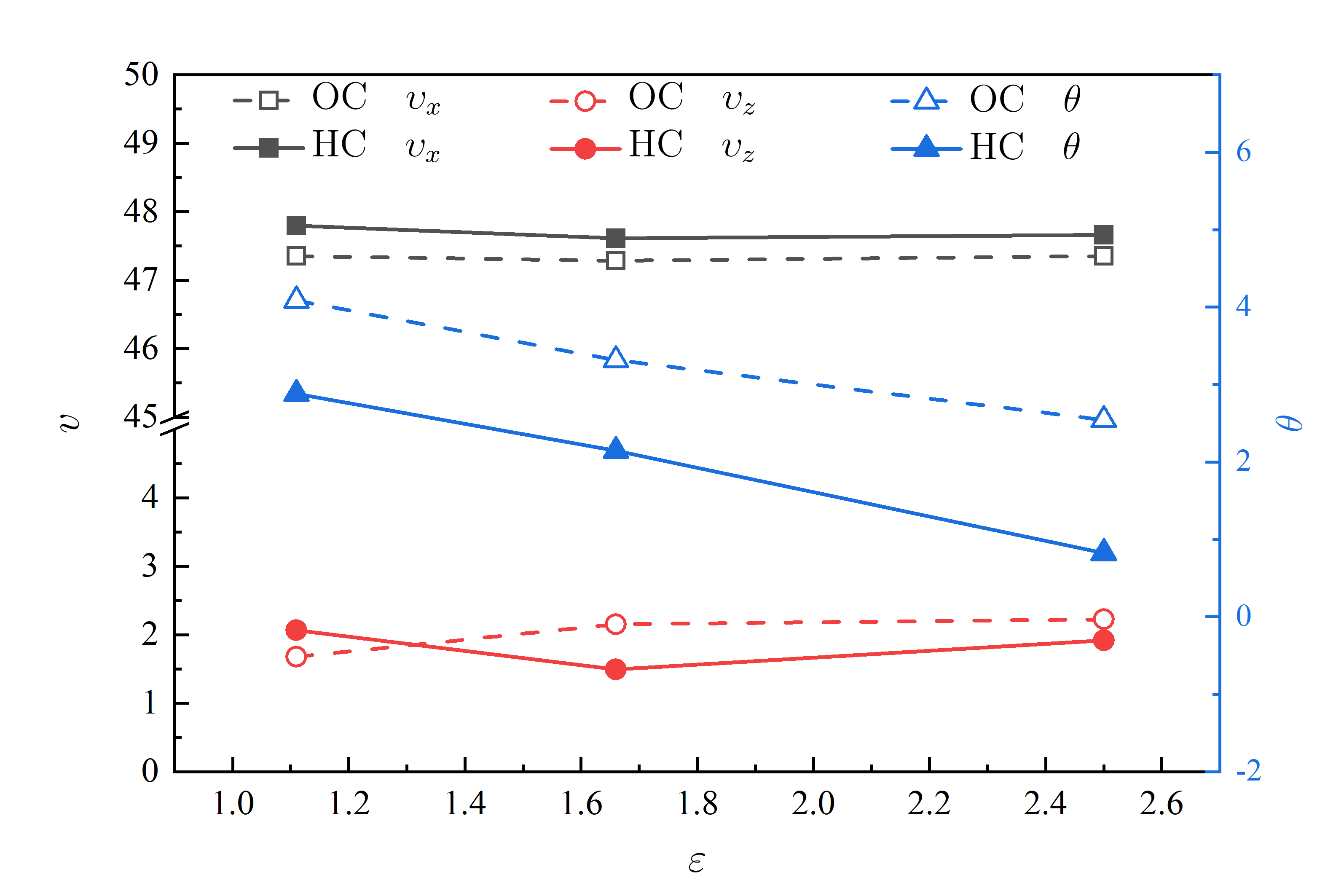}
\caption{Comparison of kinematic parameters $\upsilon_x$, $\upsilon_z$ and $\theta$ when $a_z$ reaches its peak value for OC and HC cases at various dimensionless wave lengths $\varepsilon$.}
\label{fig:CompParaDifWaveLength}
\end{figure}

For cases with a hydrofoil, a similar behaviour can be observed at the first contacting stage where the hydrofoil skims the crest instead. Encountering the second wave, the hydrofoil is first impacted with the wave, which will in practice result in a hydrodynamic lift reducing velocities and elevating the aircraft relative to the still water level, as depicted in Fig.~\ref{fig:WaveLengthCompAcc} (a) $t=$0.60s versus (d) $t=$0.75s. Furthermore, three kinematic parameters, $\upsilon_x$, $\upsilon_z$ and $\theta$, are displayed in Fig.~\ref{fig:CompParaDifWaveLength} when $a_z$ reaches its peak value for OC and HC cases. In general, the values of $\theta$ in HC cases is smaller than that in OC cases, which could reduce the area of impacting region when fuselage hits the rising wave. Moving to the comparison of vertical velocity, the results with a hydrofoil is lower than the results without a hydrofoil that can contribute a decline on $a_z$ to some extent when the impacting position is similar. For the particular case $\varepsilon=1.11$, $\upsilon_z$ of HC is greater than that of OC, whereas the load reduction on $a_z$ is still exist, that is because the relative velocity at crest is small \citep{zhang2023numerical}.
Comparing the first magenta dashed rectangle for all six cases in Fig.~\ref{fig:ContourDifWaveLength}, it can be observed that the motions are similar in character but the magnitudes and sizes of high-pressure region are much smaller for the model with hydrofoil than for the model without hydrofoil, which are responsible for the reduction of accelerations.

\subsubsection{Effect of Wave Steepness}
\begin{figure}[hbt!]
\centering
\includegraphics[width=0.6 \textwidth]{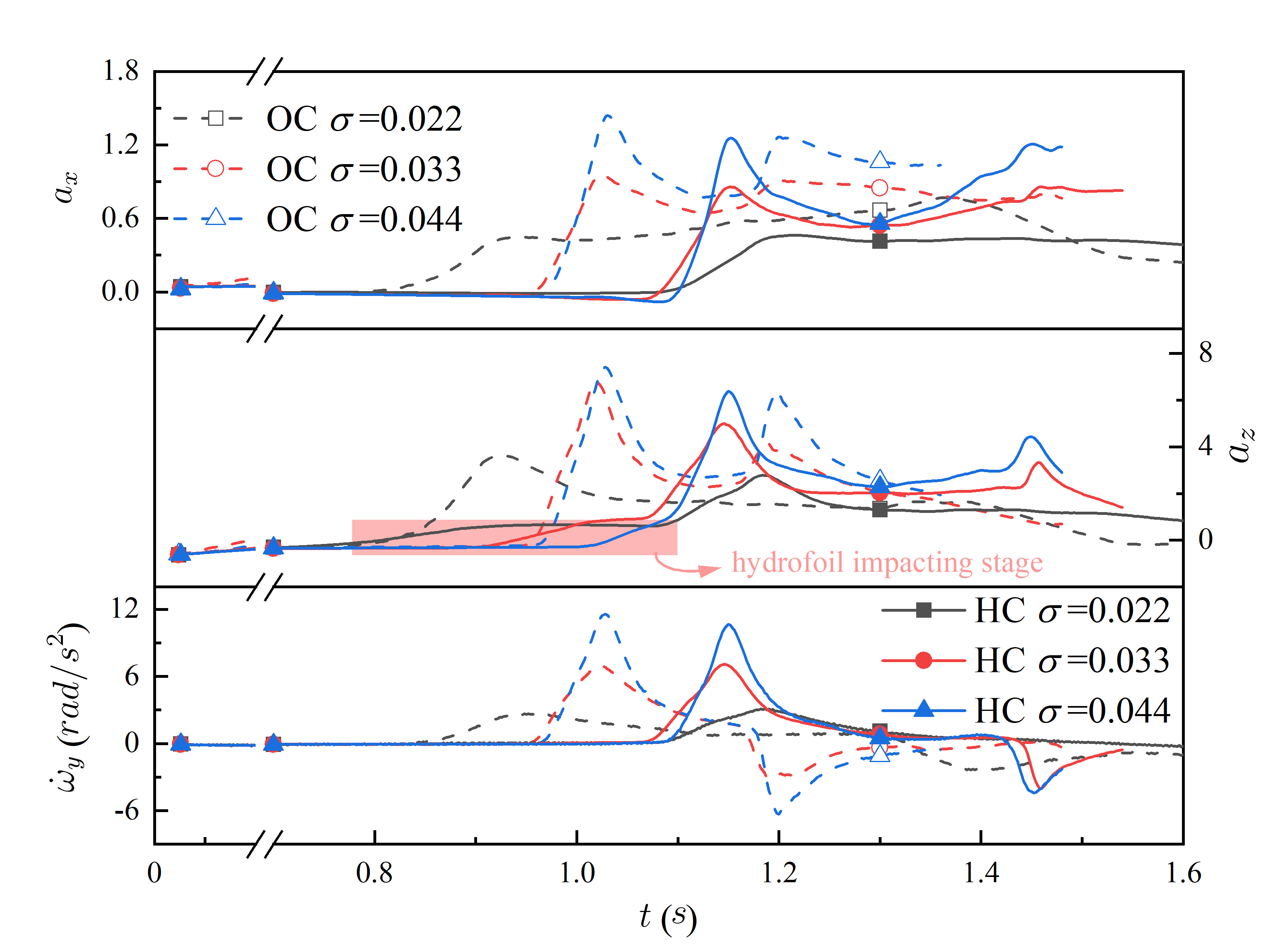}
\caption{Time histories of accelerations for original and hydrofoil configurations (OC and HC) for different dimensionless wave steepness $\sigma$.}
\label{fig:WaveHeightCompAcc}
\end{figure}

\begin{figure}[hbt!]
\centering
\includegraphics[width=0.49 \textwidth]{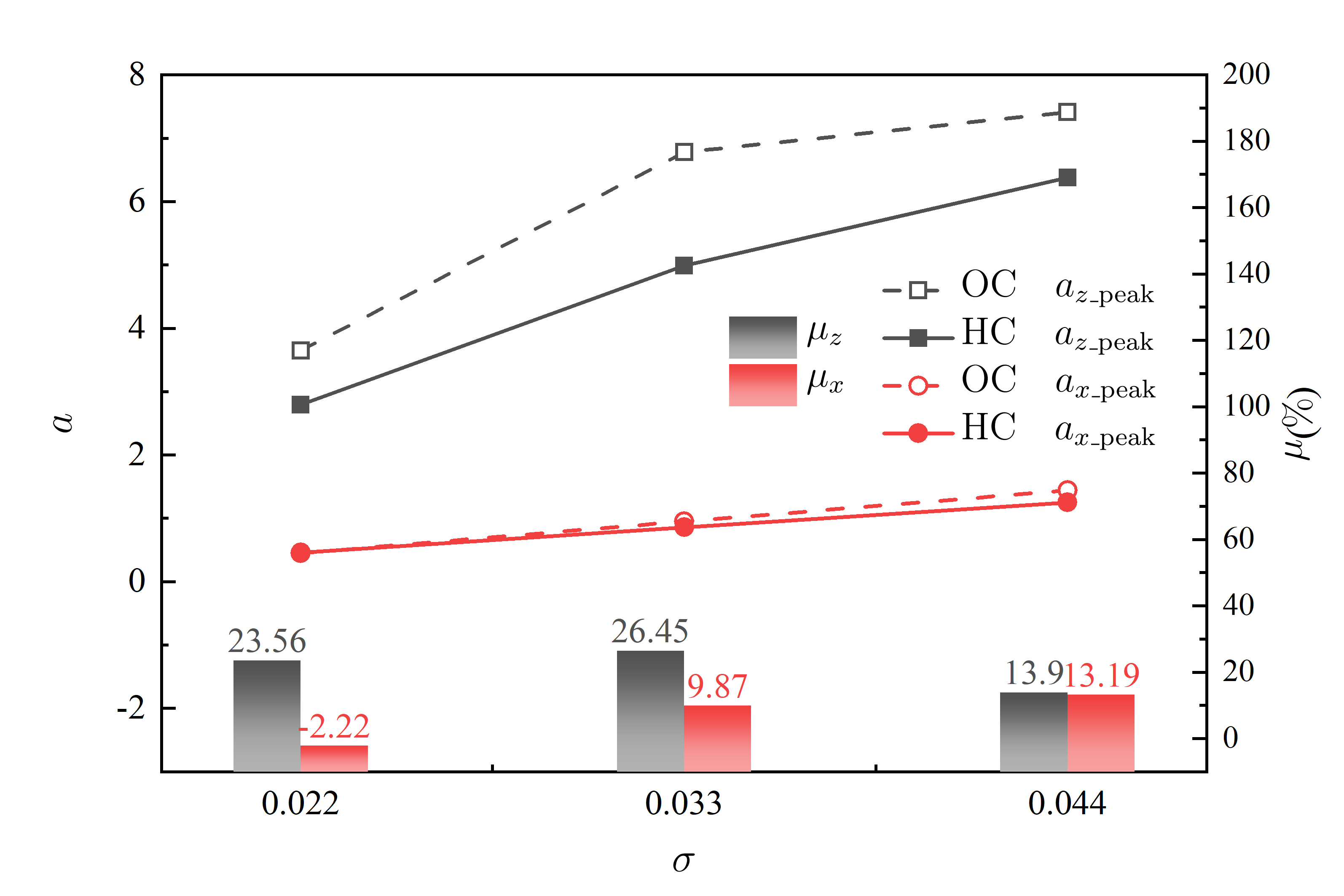}
\caption{Comparison of maximal accelerations in $z-$ and $x-$direction for OC and HC cases landing on crest with different wave steepness $\sigma$.}
\label{fig:WaveHeightCompPeak}
\end{figure}

In this section, the effect of wave height $H$ on hydrodynamic characteristics of aircraft associated with load reduction performance is discussed. Referring to the experimental study carried out by \cite{huang2019seakeeping}, the dynamic behaviour of the amphibious aircraft will encounter extreme dynamic responses when the wave length is $2\sim4$ times the length of fuselage ($2<\varepsilon<4$). From this point, the case of dimensionless wave length $\varepsilon=2.50$ is taken into consideration herein to evaluate the effect of wave height and load reducing rate, associated with three different values of wave steepness $\sigma$ ($\sigma=H/\lambda_\mathrm{w}$), i. e., 0.022, 0.033 and 0.044.
Fig.~\ref{fig:WaveHeightCompAcc} shows time histories of accelerations ($a_x$, $a_y$ and $\dot{\omega}_y$) for OC and HC cases with different $\sigma$ and the maximal values of $a_x$ and $a_z$ are displayed in Fig.~\ref{fig:WaveHeightCompPeak}. It can be observed that the accelerations become larger with increasing $\sigma$ for both OC and HC cases, indicating that the growing wave energy could be responsible for this behaviour. Moving to the comparison of maximal accelerations, the results of load reducing rate $\mu_z$ are always above $10\%$, particularly $\mu_z$ reaching up to $26.45\%$ for the case of $\sigma=0.033$. For the horizontal acceleration $a_x$, the value of $\mu_x$ is able to beyond $10\%$ with a high wave steepness.

\begin{figure}[hbt!]
\centering
\includegraphics[width=0.9\textwidth]{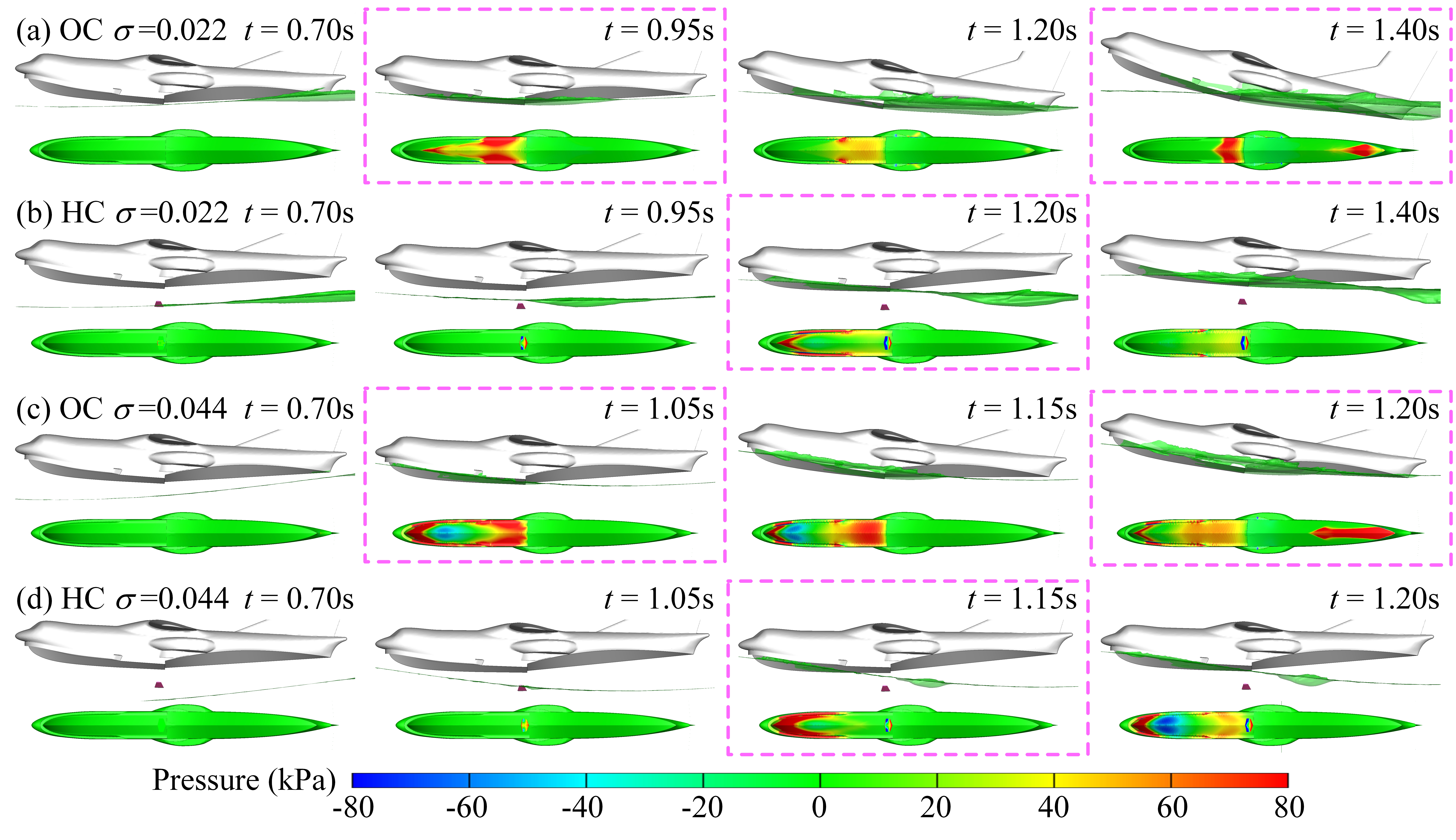}
\caption{Landing motions and pressure distribution for four time instants around the maximal vertical acceleration for different wave steepness $\sigma$.}
\label{fig:ContourDifWaveHeight}
\end{figure}

\begin{figure}[hbt!]
\centering
\includegraphics[width=0.5\textwidth]{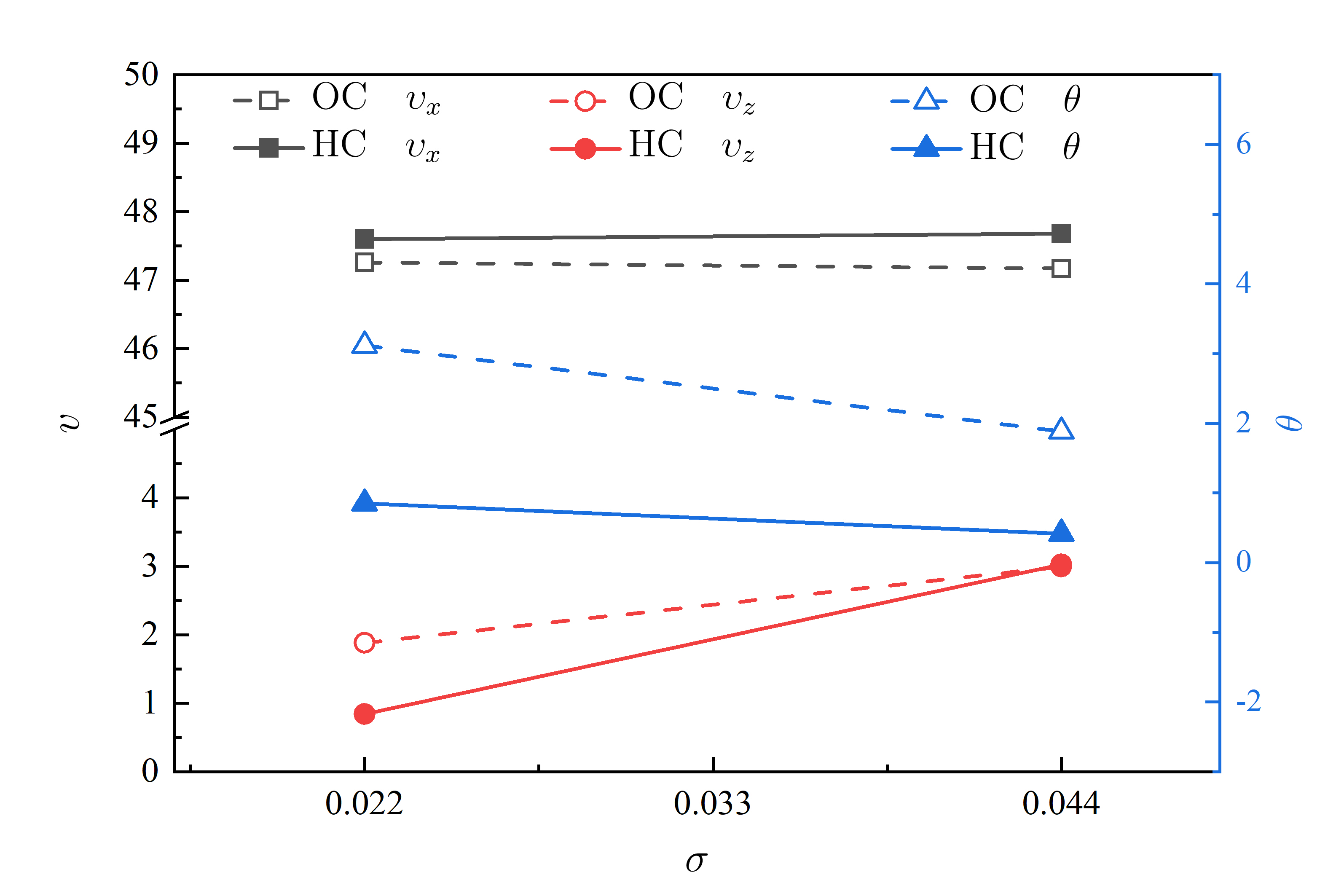}
\caption{Comparison of kinematic parameters $\upsilon_x$, $\upsilon_z$ and $\theta$ when $a_z$ reaches its peak value for OC and HC cases at various wave steepness $\sigma$.}
\label{fig:CompParaDifWaveHeight}
\end{figure}

The wave surface deformation and the pressure distribution at the bottom of fuselage are plotted and summarized in Fig.~\ref{fig:ContourDifWaveHeight} for OC and HC cases with $\sigma$=0.022 and $\sigma$=0.044. Comparing OC cases, it can be seen that for the case with a large value of $\sigma$, the body skims the first wave and then flies above the wave until impacting at the rising position of the second wave. Conversely, the aircraft in the small $\sigma$ case remains in contact with the wavy water all time and reaches its peak loading at trough. As mentioned in the work of Zhao et al. \cite{zhao2020numerical} and Zhang et al. \cite{zhang2023numerical}, the relative vertical velocity between the aircraft and wave increases from trough to rising position, and the wave particle velocities increase with the wave height as presented in Eq~\ref{eq:WaveVelocity}. Moreover, it can be observed in Fig.~\ref{fig:CompParaDifWaveHeight} that the vertical impact velocity of the aircraft also increases for larger $\sigma$. Considering the different impacting wavy position and wave energy, accelerations of OC $\sigma=0.044$ are much greater than that of OC $\sigma=0.022$, which is consistent with the pressure distribution in Fig.~\ref{fig:ContourDifWaveHeight} (a)$t=0.95$s and (c)$t=1.05$s, and similar behaviours are observed for HC cases.

Comparing OC and HC cases together, owing to the additional hydrodynamic force generated by the hydrofoil, an obvious pitch-down can be observed in Fig.~\ref{fig:CompParaDifWaveLength} when the first maximal acceleration is reached. For OC and HC cases with $\sigma=0.022$, a smaller vertical impacted velocity $\upsilon_z$ and pitching angle $\theta$ contribute to a slighter high-pressure region at the bottom of fuselage (see Fig.~\ref{fig:ContourDifWaveHeight} (a)$t=0.95$s and (d)$t=1.20$s), which brings the reduction on $a_z$. However, a minor difference can be observed on $\upsilon_z$ for case $\sigma=0.044$. At this moment, the value of pitching angle plays a second significant role on load determining the size of impacting region when encountering a similar impacting wave position. As it is shown in Fig.~\ref{fig:ContourDifWaveHeight} (c)$t=1.05$s and (d)$t=1.15$s, the pressure distribution at the bottom surface is improved.

\section{Conclusion}
\label{sec:conclusion}

In the present study, the load characteristics of the amphibious aircraft landing on still/wavy water have been investigated numerically, including the load reduction performance by using hydrofoil. Contributions and findings can be summarized as:

(1) A preliminary study on design parameters of hydrofoil, i.e., static load coefficient, sweep angle and dihedral angle, on load reducing efficiency is carried out. The static load coefficient $C_{\Delta0}$ around 24.5 is recommended due to moderate change on accelerations for the investigated range $[0.9, 196.3]$ and configurations, whereas a marginal effect of the sweep angle on accelerations is observed. Moreover, the varying range of dihedral angle $\psi$ is determined by the height of the struts to some extent, beyond which a sharp increase on hydrodynamic force will occur.

(2) For landing on still water, the initial vertical velocity plays a important role on the reduction of accelerations. A decreasing trend can be observed with decreasing $\upsilon_{z0}$ for OC and HC cases. It is worth noting that the result from the case of HC with $\upsilon_{z0}=3$ m/s is still less than that from the case of OC with $\upsilon_{z0}=1$ m/s, indicating that a significant contribution of hydrofoil to the reduction on accelerations, which is better than varying initial velocity.

(3) Looking into the wavy water conditions, since the only control applied is to keep similar initial pitching angle and initial contact position, the subsequent motion on the wave is not under control. This could cause variations in the ensuing behaviours which further lead to variations in the accelerations exerted on the aircraft. In general, the amphibious aircraft first encounters the crest wave and then hits trough or rising wave of the following wave. With the help of hydrofoil, the impacting velocity and pitching angle are reduced that is beneficial for obtaining smaller hydrodynamic force. Besides, the wave elevation velocity also needs to be considered together which affects the relative impacting velocity between the body and water surface. Specifically, for the investigated dimensionless wave length ($\varepsilon$=1.11, 1.66 and 2.50) with 3m wave height, load reduction rate $\mu_z$ in terms of $a_z$ is always greater than 25$\%$, while at least 10$\%$ can be obtained for wave steepness $\sigma$ varying from 0.022 to 0.044.


\section*{Acknowledgments}

This work has been supported by the National Natural Science Foundation of China (No. 11602200). The supports from Open Foundations of EDL Laboratory (EDL19092111) and Postgraduate Research and Practice Innovation Program of Jiangsu Province (SJCX22-0094) are also acknowledged.

\bibliography{sample}

\end{document}